\documentclass[useAMS,usenatbib]{mn2e} 
\usepackage{aas_macros}
\usepackage{graphics}
\usepackage[pdftex]{graphicx}
\usepackage{epstopdf}
\usepackage{epsfig}  
\usepackage{natbib} 
\usepackage{float}
\usepackage{amsmath}
\usepackage{times}
\usepackage[varg]{txfonts}
\usepackage{verbatim} 
\usepackage{multirow,bigdelim} 
\usepackage{color}

\title[Giant disk galaxies]{Giant disk galaxies : Where environment trumps mass in galaxy evolution}

\author[H. Courtois et al.]
{
H.M. Courtois$^{1}$\thanks{E-mail: \texttt{h.courtois@ipnl.in2p3.fr}}, 
D. Zaritsky,$^2$, J.G. Sorce$^1$ and D. Pomar\`ede$^3$
\\
$^{1}$Universit\'e Lyon 1, CNRS/IN2P3, Institut de Physique Nucl\'eaire, Lyon, France\\
$^{2}$Steward Observatory, University of Arizona, Tucson, AZ, 85721, USA\\
$^{3}$ CEA/IRFU, Saclay, 91191 Gif-sur-Yvette, France
}

\begin{document}

\date{February 2015}

\pagerange{\pageref{firstpage}--\pageref{lastpage}} \pubyear{2015}

\maketitle

\label{firstpage}

\begin{abstract}
\indent{We identify some of the most H{\small I} massive and fastest rotating disk galaxies in the local universe with the aim of 
probing the processes that drive the formation of these extreme disk galaxies. By combining
data from the Cosmic Flows project, which has consistently reanalyzed archival galaxy H{\small I}
profiles, and 3.6$\mu$m photometry obtained with the {\sl Spitzer Space Telescope}, with which we can measure stellar mass,
we use the baryonic Tully-Fisher (BTF) relationship to explore whether these massive galaxies are distinct.
We discuss several results, but the most striking is the systematic offset of the H{\small I}-massive sample above the BTF. These galaxies have both more gas and more stars in their disks than the typical disk galaxy of similar rotational velocity. The ``condensed" baryon fraction, $f_C$, the fraction of the baryons in a dark matter halo that settle either as cold gas or stars into the disk, is twice as high in the H{\small I}-massive sample than typical, and almost reaches the universal baryon fraction in some cases, suggesting that the most extreme of these galaxies have little in the way of a hot baryonic component or cold baryons distributed well outside the disk. In contrast, the star formation efficiency, measured as the ratio of the mass in stars to that in both stars and gas, shows no difference between the H{\small I}-massive sample and the typical disk galaxies.  We conclude that the star formation efficiency is driven by an internal, self-regulating process, while $f_C$ is affected by external factors. Neither the morphology nor the star formation rate of these galaxies is primarily determined by either their dark or stellar mass. We also found that the most massive HI detected galaxies are located preferentially in filaments. We present the first evidence of an environmental effect on galaxy evolution using a dynamical definition of a filament.
}

\end{abstract}



\section{Introduction}

How galaxies form and evolve remain open questions. Models
reproduce global properties of galaxies well, partly by construction,
but discriminative tests of plausible models are few and far
between. The greatest tests of theories and models often come from 
considering the extremes of parameter space.
For example, some of the most demanding tests on the current
paradigm of structure formation come from the lowest mass galaxies, which appear to be
staggeringly underabundant \citep[the ``missing satellite problem";][]{moore,klypin}. 

Here we focus
on the opposite extreme by examining some of the most massive galaxies. Rather than considering their numbers as a population, we will
examine their internal structure. While the
least massive galaxies test certain aspects of the standard paradigm,
namely the low mass end of the halo mass function, the nature of
parent vs. satellite galaxy, and the role of reionization, giant galaxies
help test the effects and results of a Hubble time of accretion, merging,
collisions, and cannibalisms of dwarf companions \citep[cf.][]{wang}.

In a theoretical framework of structure growth whose principal
characteristic is hierarchical accretion, it is natural to suspect that
the environment in which a galaxy resides plays a prominent role in
its evolution. However, the success of models in which observable
galaxy properties (luminosity, color, morphology) are based
primarily on the galaxy's stellar or dark matter mass \citep[see for examples the large literature on halo occupancy models, 
some examples include][]{peacock,berlind,conroy} has been interpreted as supporting the view
that environment plays a subdominant
role in galaxy evolution \citep{kauffmann,bundy}. 

That is not to say that environment plays no role in such models. The
parent vs. satellite nature of a galaxy \citep[cf.][]{berlind05,tinker12} is envisioned to be key
to explaining some critical observations \citep{peng,knobel}. However, disentangling the physics involved is
complicated greatly by the different degrees to which environment correlates with 
various galaxy properties, such as morphology and
color \citep{blanton}. And so, efforts to include environmental
influences in models increase the complexity and complicate interpretations of the data \citep[cf.][]{wechsler,tinker12,woo}. 

Empirically, the role of environment in galaxy evolution is often explored in the densest environments \citep[see][for a review]{boselli}, which
are necessarily the most complicated. Furthermore, the principal influence of the environment may not be direct, through phenomenon
such as ram pressure stripping or tidal interactions, but rather indirect through either ``assembly bias" \citep{sheth04, gao, wechsler}, which is purely a dark matter halo phenomenon in which close halo pairs tend to have earlier formation times than otherwise similar distant pairs, and ``history bias", in which the formation age of halos {\sl and} their 
subsequent star formation history 
varies across environments \citep{delucia}.
Identifying clean tests of the influence of mass and environment on galaxy evolution is imperative.

Defying the developed expectation that mass is universally the dominant
determinant of a galaxy's appearance and that massive galaxies should not contain cold gas 
and not be actively star forming \citep[cf.][]{dekel,cattaneo}
is a class of massive, rotation dominated
galaxy that includes H{\small I}ZOA J0836-43, the most H{\small I}-massive
galaxy known 
\cite[Sc, M$_{H{\small I}} = 7.5 \times 10^{10}$ M$_\odot$;][]{2011AJ....142..170H}, 
the fastest known rotator UGC 12591 \cite[S0/a, V$_{max} = 506$ km s$^{-1}$;][]{1986ApJ...301L...7G}, and a few other
giant disks found in the recent 40\% release of the ALFALFA survey \citep{2011AJ....142..170H}.
However, these are rare galaxies. Current determinations
of the local galaxy H{\small I}-mass function illustrate that
statistically, H{\small I}ZOA J0836-43 should not exist in the volume
explored \citep[probability of $3\times 10^{-8}$;][]{zwaan}. Nevertheless, such extreme
galaxies do exist, they are as massive as corresponding
massive early-type galaxies at low
redshift \citep{bernardi}, and they provide evidence by example against the
hegemony of mass in determining a galaxy's
properties.

Although such galaxies would clearly not be predicted in a model where mass alone determines a galaxy's properties, 
we do not know whether more sophisticated models that are currently available successfully predict the properties of such extreme objects.
Because these objects are rare, global statistical tests of the models \citep[cf.][]{obreschkow} may not recognize the absence of such galaxies.
It is therefore critical to complement large-scale comparisons, which offer 
the powerful advantage of statistics, with carefully crafted samples 
that aim to challenge the models at the margins.

These unusual galaxies must have experienced a strikingly different
evolutionary history than that of massive early type galaxies, whether that means
fewer or no large accretion events, a different angular momentum
accretion history, and/or a distinctly different star formation history. 
They are likely to be the nearest analogs of the ``classical" view of galaxy
formation \citep{eggen}, where galaxies form smoothly within an overdensity.
In fact, to produce disk-dominated galaxies numerical simulations
of spiral galaxies set in a cosmological context have had to 
select to model systems in underdense environments, a reflection
on what is now understood to relate to the details of the
angular momentum accretion history \citep[cf.][]{governato,sales}. Therefore, the importance
of environment in the genesis of these galaxies is hardwired
into detailed physics-oriented models, in opposition to what is often done in statistically-oriented
theoretical treatments.

Complementing the morphological differences between early and
late type giant galaxies, there are differences in star
formation histories. Late type giant galaxies presumably provide a
measure of the quiescent mode of star formation and a contrast to what is
likely to have occurred in early type systems. Empirically, we need
to determine whether these differences are reflected in the overall
star formation efficiency, which can be defined either relative to the total ``condensed"
baryonic content (the mass in stars and cold gas within the disk) or to the total
mass. The former measures the fraction of all ``condensed" baryons that formed stars, while the latter
measures the fraction of the total mass converted to stars and so includes any variations 
in the ``condensed" baryon fraction among galaxies.

 At this point, we pause to consider the ill-defined, and often ill-measured, description of a galaxy's mass. Theoretically, the preference would be to describe galaxies by  their total mass, where total refers to both dark and baryonic mass, or halo mass, where halo refers to the dominant dark matter component. Even there, these are moving targets because of the definition of an outer radius, such as a virial radius. However, such definitions are completely impractical in an empirical sense because we rarely have any direct measurements of an individuals galaxy's total or halo mass. Instead, investigators either default to estimating the baryonic mass, which can be problematic in its own ways (as discussed next), or to assuming that the ranking between baryonic (or, typically, more accurately stellar) mass and halo mass is nearly 1:1 \citep[see, for examples, halo abundance matching schemes;][]{berlind}. The latter approach has worked well in reproducing statistical measurements of galaxy populations from theoretical dark matter distribution models, lending support to the hypothesis that the rankings are not grossly different than 1:1. We will therefore measure baryonic masses, and presume that these are not only measurements of the baryons in these systems but also, in a ranked sense, measurements of the total mass.

The baryonic mass measurements come with their caveats as well. It is usually the case that only stellar mass, through luminosity and colors, is estimated and corrections for baryons in other phases are either ignored or treated in the mean. We will describe a somewhat more accurate process in which we use IR magnitudes that have been carefully calibrated to stellar masses and complement those with measurements of the gaseous baryons based on H{\small I} observations. This approach still ignores baryons in hot phases and there is a well-known baryon shortfall \citep{bregman}, so these masses are still not the full picture. However, to be specific, we will measure the stellar plus gaseous masses, and use these liberally as a proxy for the total galaxy mass. This assumption is broadly valid, not worse than a factor of two in total mass, as evidenced by the scatter about an apparently constant baryon fraction resulting from studies of the baryonic Tully-Fisher relation \citep{zbtf} for galaxies spanning many decades in baryon mass.

The Cosmic Flows project, which is consistently remeasuring
all available H{\small I} line widths, provides a uniform, unprecedented
dataset from which to draw these massive rotating and H{\small I}-massive
galaxies. In combination with the infrared imaging provided by the
{\sl Spitzer Space Telescope} and the calibrated conversion between 3.6
photometry and stellar mass \citep{eskew}, we are now in a position
to compare gaseous, stellar, and total masses for these rare
rotationally supported giant galaxies. In \S2 we present the data and present our two 
subsamples of extreme galaxies in \S3. In \S4 we examine the properties of these galaxies 
relative to the baryonic Tully-Fisher relation and draw inferences regarding the ``condensed" baryon fractions and star formation efficiencies. We summarize our conclusions in \S5. Where needed, we adopt H$_0 = 70$ km s$^{-1}$ Mpc$^{-1}$.

\section{The Data}

\subsection{The Cosmic Flows Program}

Since 2009, the Cosmic Flows project (CF) has gathered all the
digital H{\small I} spectra available from public archives of the largest
radio-telescopes worldwide and measured them in a consistent
way. Two sub-projects of CF, at Green Bank in the USA and at
Parkes in Australia \citep{2011MNRAS.414.2005C}, complete the archives for targets without
previous observations adequate for the Tully-Fisher method of
distance measurements \citep{1977A&amp;A....54..661T}. The latest update of this catalog
is published in  \cite{CT2015}.

The main goal of CF is to map the all-sky peculiar velocity
field locally and to uncover, at high spatial resolution, the
underlying dark matter distribution. In that cause, they have measured
tens of thousands of galaxy line widths with a new method
described by \cite{2009AJ....138.1938C} and \cite{2011MNRAS.414.2005C}. Briefly, for each galaxy, they evaluate the
line width parameter Wm50, which measures the H{\small I} profile width
at 50\% of the mean flux within the velocity range encompassing
90\% of the total H{\small I} flux. That quantity is transformed into the more
physically motivated parameter W$_{mx}$, which is the same width
corrected for the slight relativistic broadening and for broadening
due to finite spectral resolution, corrected for inclination and so is
effectively twice the maximum rotation velocity. Technical details
of this process are provided by \cite{2009AJ....138.1938C} and \cite{2011MNRAS.414.2005C}, and reviewed in \cite{2012ApJ...749...78T}.

The result is a catalog of H{\small I} measurements of unprecedented
size (14221 galaxies) and consistency. They use the ratio of the
signal level at 50\% of the mean flux to the noise measured beyond the
frequency extremities of the signal to parametrize the line width
uncertainty. Only profiles with uncertainty estimates smaller than
or equal to 20 km s$^{-1}$ are retained (more than 75\% of the 14221
galaxies, exactly 10733 galaxies satisfy this criteria) after a supplementary
visual inspection. This catalog is available for public use at
the Extragalactic Distance Database (EDD) website\footnote{http://edd.ifa.hawaii.edu; catalog ``All Digital H{\small I}"} where several
other parameters, such as the integrated H{\small I} line fluxes for those
lines with a flux calibration error better than
about 10 to 15\% and the average heliocentric velocities, are also
available and described \citep{2009AJ....138..323T}.
In addition to this ``all-digital" catalog, we also use the
``pre-digital" catalog of measurements at 20\% of the H{\small I} line peaks.
The relation
\begin{equation}
V_{m50} ({\rm all - digital}) = (1.015 \times W_{20}({\rm pre -digital}) - 11.25)
\end{equation}
connects the two catalogs \citep{2009AJ....138.1938C}. In total, we have 16,124 galaxies with
coherent H{\small I} measurements and 12,189 with sufficiently good quality
for distance measurements.

\subsection{The Infrared Data}

We use 3.6$\mu$m magnitudes measured from data obtained using the {\sl Spitzer Space Telescope} \citep{2004ApJS..154....1W} and the IRAC camera \citep{2004ApJS..154...10F}, from a number of different observing programs \citep[such as,][]{2009ApJ...703..517D,sheth,2012AJ....144..133S}. For Tully-Fisher studies, within the CF sample, there are 6,007 galaxies with sufficiently good H{\small I} linewidth measurements, $\sigma_{Wm50} < 20$ km s$^{-1}$ and inclinations greater than 45$^\circ$. Among these, 2,493 have been observed with {\sl Spitzer}. We adopt total magnitudes measured either using the Archangel pipeline \citep{2007astro.ph..3646S} or the S$^4$G pipeline \citep{munoz}.
Distances, needed to calculated absolute magnitudes, are presumed from the Hubble flow using H$_0 = $ 70 km s$^{-1}$ Mpc$^{-1}$ and the recessional velocities from the H{\small I} catalog.

We use the 3.6$\mu$m photometry because there now exist simple, and robust, conversions to stellar mass \citep{eskew,meidt14} that have been vetted with comparisons to SDSS SED-derived stellar masses \citep{cybulski}. \citet{eskew} give the conversion between stellar mass and infrared flux as M$_* = 10^{5.97} F_{3.6} ({D\over 0.05})^2$, where D is the distance to the galaxy in Mpc, $F_{3.6}$ is the flux in Jy, and M$_*$ is expressed in solar masses. A slightly more precise estimate can be obtained by including the 4.5$\mu$m flux, but both \citet{eskew} and \citet{querejeta} find only a weak dependence between M$_*$ and infrared color. The largest uncertainty in the derived stellar mass remains the overall normalization due to uncertainties in the adoption of a specific stellar initial mass function (the \cite{eskew} calculation adopts a Salpeter IMF). 

\section{GIANT DISK GALAXIES}

To define our galaxy subsamples, we search both for those galaxies with 
the largest H{\small I} masses and for those
with the largest rotational velocities. Due to our use of the H{\small I} database, we may miss gas-poor versions of the latter, although such galaxies are likely to be early type and therefore not the pristine test cases we are searching for. 
Because of broad correlation between gas mass and total mass, these samples have some objects in common.

\subsection{The most H{\small I} massive galaxies}

We compute H{\small I} masses (M$_{\rm HI}$)
from the publicly available integral fluxes and average heliocentric
velocities \citep{2009AJ....138.1938C} using

\begin{equation}
{\rm M}_{\rm HI}= 2.36 \times 10^5 {\rm D}^2 {\rm FI}
\end{equation}
\noindent
where D is the distance in Mpc and FI is the integrated flux in the H{\small I}
line in Jy km s$^{-1}$. 

\begin{figure}
\centering
\includegraphics[scale=0.4]{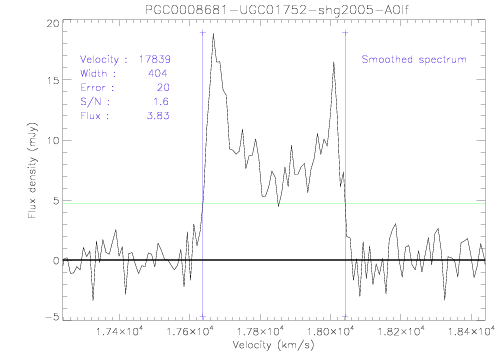}
\includegraphics[scale=0.5]{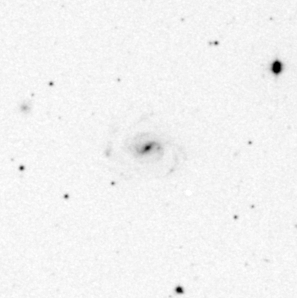}
\caption{H{\small I} profile of PGC0008681 as observed at Arecibo, and B band image from DSS2 survey at the size of Arecibo beam (3 arcmin). This is the largest H{\small I} mass galaxy recorded in the full available sample, but is not in our subsample due to lack of {\sl Spitzer} photometry.}
\label{fig:PGC0008681}
\end{figure}

\begin{figure}
\centering
\includegraphics[scale=0.5]{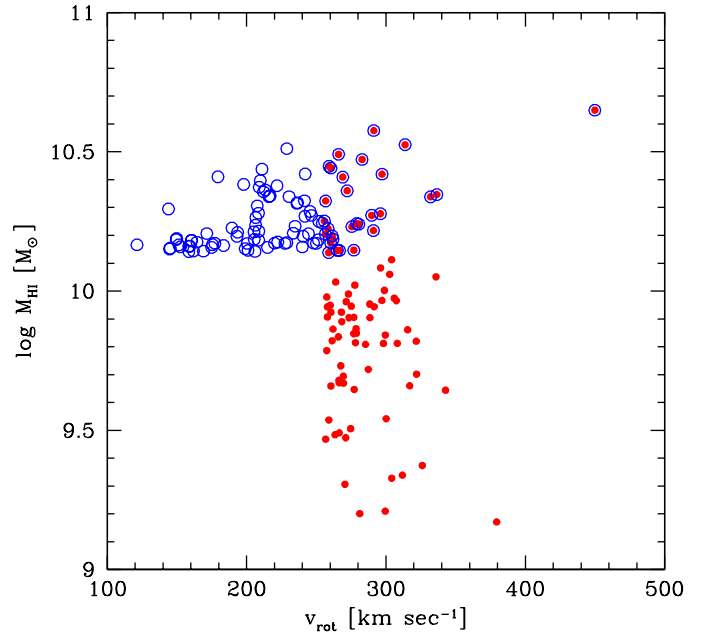}
\caption{Distribution of galaxies in our subsamples selected by rotational velocity and H{\small I} mass. We highlight the H{\small I} mass selected sample using open blue circles and the rotational velocity selected sample using filled red circles. We will continue this color scheme throughout the paper to differentiate the samples. There are 26 galaxies in common between the two samples.}
\label{fig:sample}
\end{figure}

The distribution in log M$_{\rm HI}$ for the 10733 galaxies for which we can calculate the necessary quantities covers the range of 5.53 to 10.72, with a median of 
9.62, corresponding to a range of $3.4\times 10^5$ to $5.2\times 10^{10}$M$_\odot$. PGC8681/UGC01752 is the most H{\small I} massive
(Figure \ref{fig:PGC0008681}) galaxy in the full sample, but it is not in our subsample because we lack {\sl Spitzer} photometry for it. PGC17625, with only a slightly lower H{\small I} mass ($4.5\times10^{10} {\rm M}_\odot$), tops our list. In Figure \ref{fig:sample} we show the distribution of M$_{\rm HI}$ for the 100 most H{\small I} massive galaxies that also have existing {\sl Spitzer} photometry. Choosing the top 100, which was set only because it is a simple round number, corresponds roughly to selecting the top 5\% of the available sample. One galaxy (PGC90167) was removed from our subsample because even though it was observed with {\sl Spitzer}, the S$^4$G survey \citep{sheth,munoz} was unable to provide a reliable magnitude measurement. 

We distinguish this sample from an H{\small I} rich sample, where in that case one would presumably select on either the ratio of gas to stellar mass, M$_{\rm HI}$/M$_*$,  or to dark matter mass, M$_{\rm HI}$/M$_{\rm DM}$. Although such a sample is also of interest, that selection depends on the modeling of M$_*$ or M$_{\rm DM}$ and therefore complicates any interpretation.

\subsection{The fastest rotators}

Our rotational velocities come from the $W_{mx}$ parameter:

\begin{equation}
{\rm v}_{rot}=\frac{W_{mx}/2}{\sin (i)}
\end{equation}
where v$_{rot}$ is the galaxy rotational velocity and $i$ the inclination to the line of sight.

\begin{figure}
\centering
\includegraphics[scale=0.4]{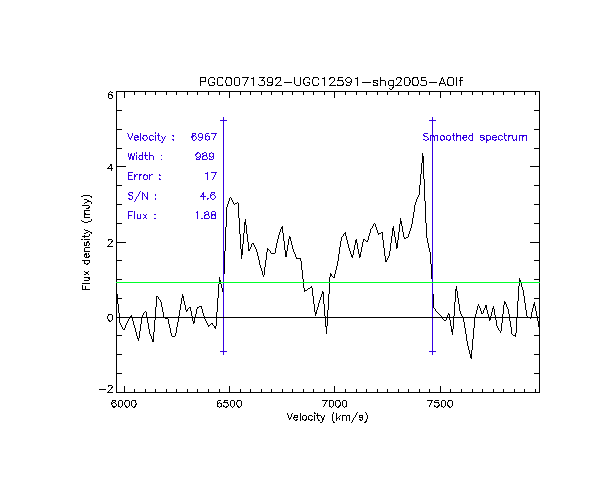}
\includegraphics[scale=0.5]{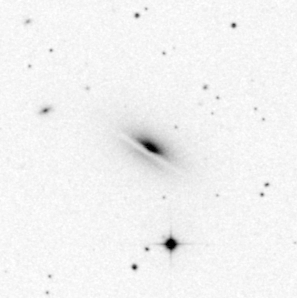}
\caption{H{\small I} profile of PGC0071392 as observed at Arecibo, and B band image from DSS2 at the size of Arecibo beam (3 arcmin). This is the largest rotational velocity recorded in the full available sample, but is not in our subsample due to lack of {\sl Spitzer} photometry.}
\label{fig:PGC0071392}
\end{figure}

Among the full sample of galaxies (not yet implementing the {\sl Spitzer} imaging requirement), the rotational velocities range from 11 to 514 km s$^{-1}$ for PGC71392/UGC12591 (see Figure \ref{fig:PGC0071392}) with a median of 145 km s$^{-1}$. Again the top ranked galaxy is not in our final sample due to the lack of infrared photometry, and so the top ranked galaxy in our sample is PGC17625 with a rotation velocity of 450 km s$^{-1}$. The distribution of rotational velocities of our sample is also illustrated in Figure \ref{fig:sample}.

\begin{figure}
\centering
\includegraphics[scale=0.4]{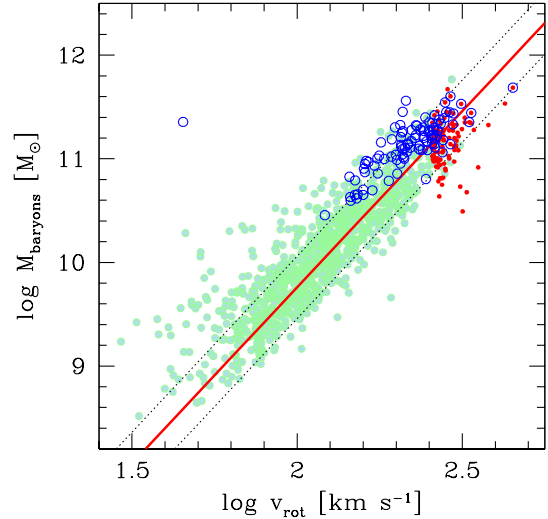}
\caption{ 
Baryonic Tully Fisher. Comparison of the BTF for the S$^4$G sample and our subsamples of extreme disks. 
The red filled circles represent the fast rotators, while the blue open circles represent the H{\small I}-massive galaxies. The red solid line is described in detail by \citet{zbtf} but in brief represents a model where a fixed fraction ($\sim$ 0.4) of the baryons in a halo of circular velocity of v$_{rot}$, which has the cosmological baryon fraction, are settled in the disk. As shown by those authors, this simple model is consistent with the empirical best fit line.
The dotted lines represent factors of two change in that fixed fraction.}
\label{fig:btf}
\end{figure}

\section{Discussion}

\subsection{The baryonic Tully-Fisher relationship}

Because in any sample there will always be some galaxies with the largest H{\small I} masses or rotation velocities, we require a fiducial against which to compare these galaxies to the average galaxies. Are they simply scaled up versions of lesser galaxies or are they otherwise distinct?

This class of question can be addressed using galaxy scaling relations. For rotationally supported galaxies the most commonly used such relation is that identified by \citet{1977A&amp;A....54..661T}, which relates rotational velocity to optical luminosity. One can then use such a relation to ask such questions as whether these galaxies have unusual luminosites for their rotational velocities. 

Since the original work, this scaling relation has been extended in one key way that has been shown to be particularly necessary for lower mass galaxies, namely the substitution of the baryonic mass in place of the luminosity \citep{freeman,walker,mcgaugh00,verheijen,geha}. For our purposes, we will use the baryonic Tully-Fisher (BTF) measured for S$^4$G galaxies \citep{zbtf} as our fiducial because the data used to establish that relation are most comparable to the data presented here.

To implement the BTF we require measurements of the stellar and gaseous masses.  We described in \S2.2 how we obtain M$_*$. Obtaining a measurement of the gaseous mass, M$_{\rm GAS}$, involves correcting the H{\small I} mass for the missing He, other ``metals", and molecular gas. This is done here as described by \cite{zbtf} to enable a direct comparison to their results. The baryonic mass is then simply ${\rm M}_* +  {\rm M}_{\rm GAS}$. As noted by those authors, this quantity is more accurately referred to as the ``condensed" baryonic mass because it only represents the baryons that have settled into the disk of the galaxy and is missing the possibly substantial baryonic components in the halos of these galaxies.

\begin{figure}
\centering
\includegraphics[scale=0.4]{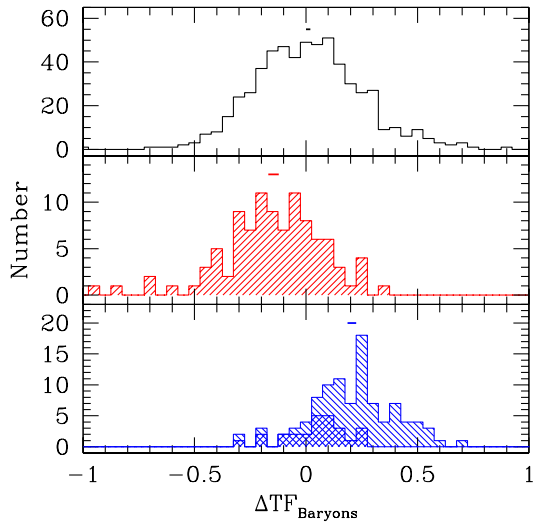}
\caption{Deviations from the mean BTF relation, $\Delta {\rm TF}_{Baryon}$, for the BTF shown in Figure \ref{fig:btf}, for galaxies with v$_{rot} > 2$. Upper panel shows the residual distribution for the S$^4$G sample from \citet{zbtf}, middle panel for our subsample of fast rotators, and lower panel for our subsample of H{\small I}-massive galaxies. Also shown in the lower panel, with the opposite hashing is the subpopulation of H{\small I}-massive galaxies that are also in the fast-rotator sample. The small horizontal bars placed above each histogram represent the mean of each distribution and the length represents the standard deviation of that mean.}
\label{fig:hist}
\end{figure}

In Figure \ref{fig:btf} we present our first comparison of the properties of the giant galaxies to the more general population. As shown, both the fastest rotators and most H{\small I} massive galaxies lie generally on the BTF relation, although some differences exist. First, although slightly difficult to notice from this Figure but easier to spot in Figure \ref{fig:hist} where we plot the residuals from the mean BTF ($\Delta$TF$_{\rm Baryons}$), the fastest rotators fall somewhat below the mean trend as depicted by the red line in Figure \ref{fig:btf}. Second, and much clearer, the H{\small I}-massive sample lies systematically above the line. The one striking outlier,
PGC90167, has a T-Type $= -2$.  Because it is such an early-type galaxy, its dynamical support is probably not dominated by rotational support and therefore the galaxy should not be expected to satisfy the BTF (it has a low v$_{rot}$ of 45 km s$^{-1}$). 
We remove this one galaxy from further consideration, leaving both the fast rotator and H{\small I}-massive samples with 99 galaxies, and proceed to discuss the two subsamples and their location relative to the mean BTF relation. 

There are a variety of potential explanations for why the fast rotators fall below the mean BTF:

\smallskip
\noindent
First, this result could be a manifestation of a Malmquist-like bias. Because we select  this subsample using v$_{rot}$, errors that artificially inflate v$_{rot}$ will preferentially contaminate our sample with galaxies that appear to be ``baryon poor" for their rotation velocity. The mean of the distribution (Figure \ref{fig:hist}) is displaced by $\sim$ 0.15 dex, suggesting that we would need to have a similar magnitude error in v$_{rot}$ for this to be a plausible explanation. That level of error translates to an error of $\sim$ 40\% or over 100 km s$^{-1}$ for v$_{rot} = 300$ km s$^{-1}$. Given our demands on precisely determined H{\small I} widths and high inclinations, this level of error seems unlikely. 

\smallskip
\noindent
Second, inclination errors, in the form of an underestimation of the true inclination, could lead us to preferentially place objects to the right of the mean BTF. We test for this effect by searching for a correlation between inclination and $\Delta$TF$_{\rm Baryons}$. There is only a very weak positive correlation (positive is in the necessary sense) that is entirely consistent with arising at random (can be rejected as arising randomly with only 37\% confidence). 
We conclude there is no evidence that inclination is systematically affecting the BTF residuals of the high v$_{rot}$ galaxies.

\smallskip
\noindent
Third, at these high v$_{rot}$ values we may have a larger fraction of galaxies for which pressure supported components play a larger dynamical role. \citet{2007MNRAS.381.1463N}, when examining the high mass end of the Tully-Fisher (TF) relation, found that a number of their galaxies were S0/Sa types (as well as finding a similar displacement of these massive galaxies below the general TF). When dealing with such galaxies a number of investigators \citep{burstein,weiner,kassin,zall} have suggested combining the use of v$_{rot}$ and velocity dispersion, $\sigma$, into a single kinematic term that measures the full dynamical support. While this may be appropriate when observing stellar components, it is not for gaseous measurements (there is negligible pressure support of the gas component and, even if there were, such a correction would move these galaxies even further to the right in the Figure). If one does attribute the dip below the BTF for this subsample to the prevalence of early-type galaxies at these rotation velocities, the cause of the dip would have to be a general failure of (or departure from) the linear BTF, rather than from the omission of $\sigma$ in the evaluation of the dynamical support. 

\smallskip
\noindent
Fourth, the intrinsic BTF may deviate from a straight line. Numerical simulations \citep{aumer} also find that the most massive galaxies tail off below the BTF \citep{zbtf} and so this phenomenon may be a natural result, for which an intuitive explanation might provide significant insight to the nature of galaxy formation. As a potentially interesting aside, massive early type galaxies also appear to fall below their analogous scaling relation, the Fundamental Plane \citep{bernardi}. As for the early types, for which curvature in the scaling relation is established \citep{zfm,bernardi}, a unified scaling relation \citep{zall,zrev} would imply the same phenomenon for late types.

\smallskip
\noindent
Finally, the position of these galaxies may indicate that the stellar mass has been underestimated in these systems relative to that of the lower v$_{rot}$ systems. 
This error could arise because the stellar initial mass function (IMF) is more bottom heavy in the faster rotators. Although this hypothesis may seem farfetched,
there is empirical evidence for this type of behavior among early type galaxies \citep{vandokkum,cappellari}, where the IMF becomes progressively more bottom heavy as one considers galaxies with larger velocity dispersions. Evidence for IMF variations even extends to Local Group clusters \citep{z12,z14}, and so the possibility that such variations exist among disk galaxies as well should not be quickly dismissed.

\begin{figure}
\centering
\includegraphics[scale=0.4]{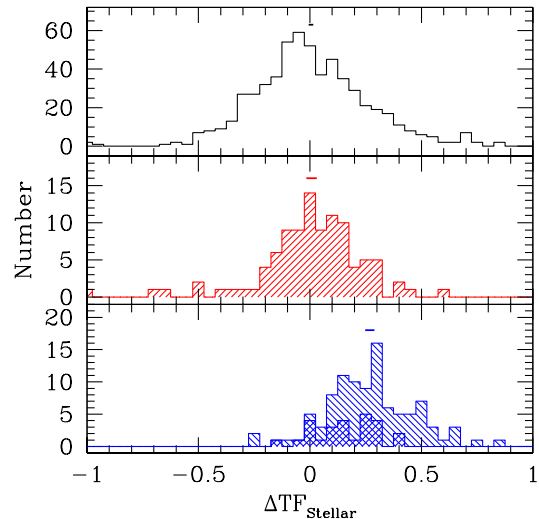}
\caption{Deviations from the stellar mass Tully-Fisher relation, $\Delta {\rm TF}_{Stellar}$, where the stellar TF is in effect the classic TF because the stellar masses are recovered directly from the luminosities. Panels and lines indicating means of the distribution are as in Figure \ref{fig:hist}. The deviation of the H{\small I} selected galaxies is as strong as in Figure \ref{fig:hist} even though these residuals are independent of H{\small I} mass.}
\label{fig:stars}
\end{figure}

As already mentioned, the more striking  departure from the BTF is seen for the H{\small I}-massive subsample. Here the departure, shown clearly in Figures \ref{fig:btf} and  \ref{fig:hist}, can be interpreted to mean that these galaxies have a larger fraction of their cosmologically apportionment of baryons within their disks, as stars or gas, than the typical disk galaxy. As with the fast rotator sample, there are a set of potential systematic errors that could also produce this result:

\smallskip
\noindent
First, one might wonder whether this could be the result of a Malmquist-like bias. We select systems with large H{\small I} masses, and therefore errors in the measured H{\small I} mass could both help populate the sample and make it appear as if these galaxies are H{\small I} rich for their rotation velocity. However, these galaxies also have larger stellar masses than typical, as shown in Figure \ref{fig:stars} where we compare to a stellar mass only version of the BTF (renormalized to produce a mean departure of zero for galaxies with log v$_{rot} > 2$ in S$^4$G). Because the stellar masses, and this stellar version of the BTF, are entirely independent of M$_{\rm HI}$, we conclude that the departure of these galaxies from the BTF is not a result of errors in the H{\small I} measurements themselves. 

\smallskip
\noindent
Second, errors in the adopted distance would affect both the gaseous and stellar masses because distance plays a role in converting between observables and masses. The mean shift between either the gaseous or stellar masses and the fiducial of the S$^4$G sample is $\sim 0.22$ dex (see Figures \ref{fig:hist} and \ref{fig:stars}). Because both mass estimates are proportional to $D^2$, this offset corresponds to a systematic offset of 0.11 dex in $D$ or alternatively a 30\% error. We know such an error cannot arise from H$_0$ given the current degree of uncertainty in the Hubble parameter \citep{freedman}. Peculiar motions could be as large as several thousand km s$^{-1}$ in the richest galaxy clusters, but are typically a few hundred km s$^{-1}$ \citep{peebles,kaiser}, particularly in the low density environments that these galaxies reside in. One might dismiss peculiar motions on the basis that these should result in increased scatter rather than an offset, but recall that we select the most  H{\small I} massive galaxies, and therefore are predisposed toward galaxies with overestimated distances. However, even with peculiar motions that are as large as 500 km s$^{-1}$, the induced error is $<$ 30\% in 97 out of the 99 galaxies in the H{\small I}-massive sample. We conclude that distance errors are not responsible for the systematic offset of H{\small I} massive galaxies from the BTF.

\smallskip
\noindent
Third, inclination errors will affect v$_{rot}$. However, unlike in the previous discussion,  inclination does not play a role in the selection of these galaxies (other than the requirement that they be inclined by more 45$^\circ$). As before, we test for the influence of inclination by searching for a correlation between measured inclination and the offset from the BTF. Again we find no significant correlation (the probability of the observed correlation arising at random is 18\%). We conclude that inclination errors are not driving the BTF residuals.

\smallskip
\noindent
Finally, the adopted relation between v$_{rot}$ and halo mass may be incorrect. We have the taken a mean trend from \cite{bullock}, normalized using the Milky Way \citep[see][]{zbtf}, but different types of galaxies will have different degrees of adiabatic contraction, and therefore different relations between the inner rotation curve, v$_{rot}$, and the characteristic
halo circular velocity, v$_c$. A factor of two overestimation in the fraction of baryons within the halo that have settled into a disk, $f_C$, would arise if we had underestimated the halo mass of galaxies in our subsample by a factor of two, or alternatively underestimated v$_{c}$ by $\sqrt{2}$. Because it is difficult to imagine a scenario by which v$_{rot}$ is smaller than that due to the dark matter, we would posit that the problem, if it exists at all, is that the v$_c$ of the typical galaxies are overestimated by a factor of $\sqrt{2}$. For a typical galaxy in our sample with v$_{rot} \sim 180$  km s$^{-1}$, the necessary error would require the halo to have $v_{rot} \sim 127$. Such a steep drop between v$_{rot}$ and  v$_c$ is not found in dynamical analyses \citep{zaritsky94} or gravitational lensing \citep{fischer}. While the conversion between v$_{rot}$ and v$_c$ has its associated systematic uncertainties, we conclude that it is unlikely to globally cause a factor of two increase in the apparent $f_C$ of the H{\small I} massive sample.

\begin{figure}
\centering
\includegraphics[scale=0.4]{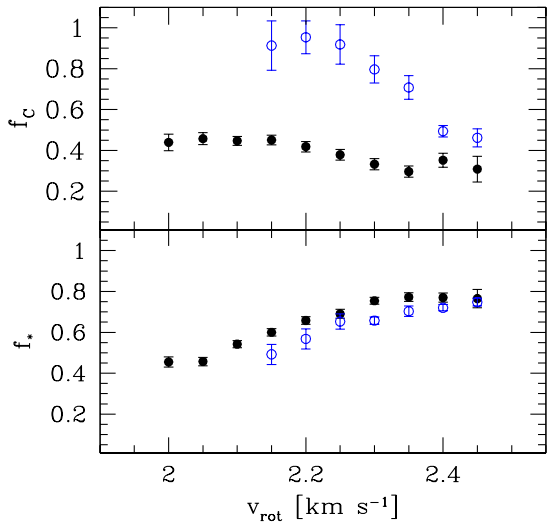}
\caption{The condensed baryon fraction, $f_C$, which is defined as the fraction of all the cosmologically expected baryons in a halo that are present in the disk as either cold gas or stars, as a function of v$_{rot}$ (upper panel) and the star formation efficiency, $f_*$, which is defined as the fraction of condensed baryons that are in stars ($M_*/(M_*+M_{GAS}))$. In this figure blue open circles represent the H{\small I} selected sample and black filled circles represent the S$^4$G sample. Points only plotted if bin has at least 3 galaxies in it. Error bars represent error in the mean.}
\label{fig:efficiency}
\end{figure}

In the upper panel of Figure \ref{fig:efficiency},  we calculate how the BTF offset translates into differences in the level of baryon ``condensation" in galaxies \citep[see][for the original discussion of this term and the calculation relative to the BTF]{zbtf} as a function of v$_{rot}$. In other words, the Figure shows the fraction of baryons --- stars and gas --- apportioned to a given halo at the level of the universal baryon fraction that settle into the disks of these galaxies.  In the lower panel of the same Figure, we plot the star formation efficiency, or alternatively the fraction of the ``condensed" mass that is turned into stars, $f_* \equiv {\rm M}_*/({\rm M}_*+{\rm M}_{GAS})$. 

The H{\small I} massive galaxies are clearly distinct from the typical disk galaxy in that the fraction of condensed baryons, which in some cases reaches nearly 100\% of the cosmological baryon fraction (meaning that all of the baryons expected in a halo of the given mass are in gas and stars within the disks), is significantly larger over nearly the full range of v$_{rot}$. 
We speculate that the reason that the difference becomes less striking at higher v$_{rot}$ is that the H{\small I} mass limit on which we select these galaxies is approaching the typical H{\small I} mass for these more massive galaxies.

In contrast, the star formation efficiency is remarkably similar for the H{\small I} massive galaxies and typical S$^4$G galaxies, even as that efficiency appears to depend on v$_{rot}$. 
The rise in $f_*$ is interesting but difficult to interpret because it probably also reflects a change in the morphological mix of galaxies.  It is nevertheless, particularly interesting that among the H{\small I} massive galaxies, the efficiency follows that of the S$^4$G galaxies despite variations of a factor of two in $f_C$. This result suggests that star formation is strongly governed by the available gas mass and internal regulation, a result that is at least broadly reminiscent of results obtained in detailed studies of star formation rates in galaxies \citep{kennicutt}, rather than external factors.

\subsection{Environment}

Given our general understanding that massive galaxies tend to be early types, the 
physically interesting question is why the massive galaxies we have selected are such strong examples of late type galaxies. If the two primary drivers of galaxy evolutionary differences among galaxies are mass and environment, and if mass is not the cause of the differences seen here, then 
environment has somehow played a dominant role for these galaxies. Confirming this conclusion is, however, difficult. By the nature of the selection of the sample (isolated so that the H{\small I} beam is clean and the result is a well-defined double horned profile), we have already selected against galaxies in dense environments. Upon examination, we confirm that none of these galaxies is in a highly overdense environment, such as an Abell cluster. 
Looking a bit further, we searched for differences in the environments of these galaxies relative to galaxies in the 2MASS redshift catalog (2MRS) \citep{2012ApJS..199...26H} by comparing the correlation function of our galaxies to that of similar galaxies in the redshift catalog. We found no convincing evidence of a difference.

We face two difficulties in exploring the role of environment further. First, environment means many different things. The local environment, whether a galaxy is a parent or satellite, appears to 
play an important role in the observable characteristics of galaxies \citep{berlind05,tinker12}. On the other hand, the large scale (several Mpc) environment also appears to play a role \citep{gomez}. Therefore, any one measure of environment will be a far from perfect diagnostic of the role of environment in evolution.  Second, galaxies, particularly most massive galaxies, have inhabited many different environments over their lifetime. The environment we measure today, especially for those in denser than average environments, is not a complete measure of the environmental history of a galaxy \citep{delucia}. Both of these issues suggest that standard analyses may at best yield only subtle differences, even if environment does play a role in a galaxy's evolution.

\begin{figure}
\centering
\hspace{-0.5cm}
\includegraphics[width=7cm]{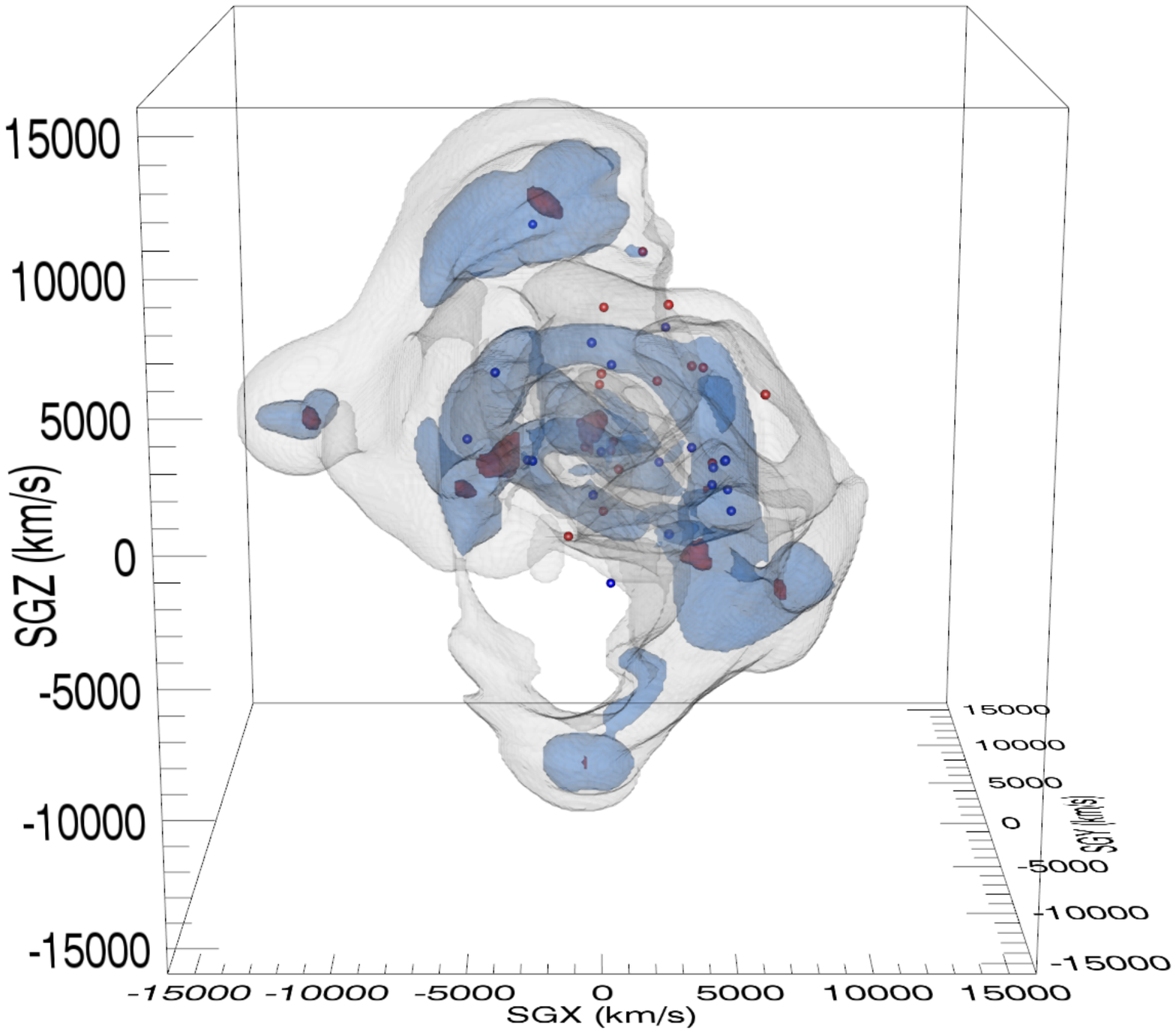}\\
\vspace{-4cm}
\includegraphics[width=7cm]{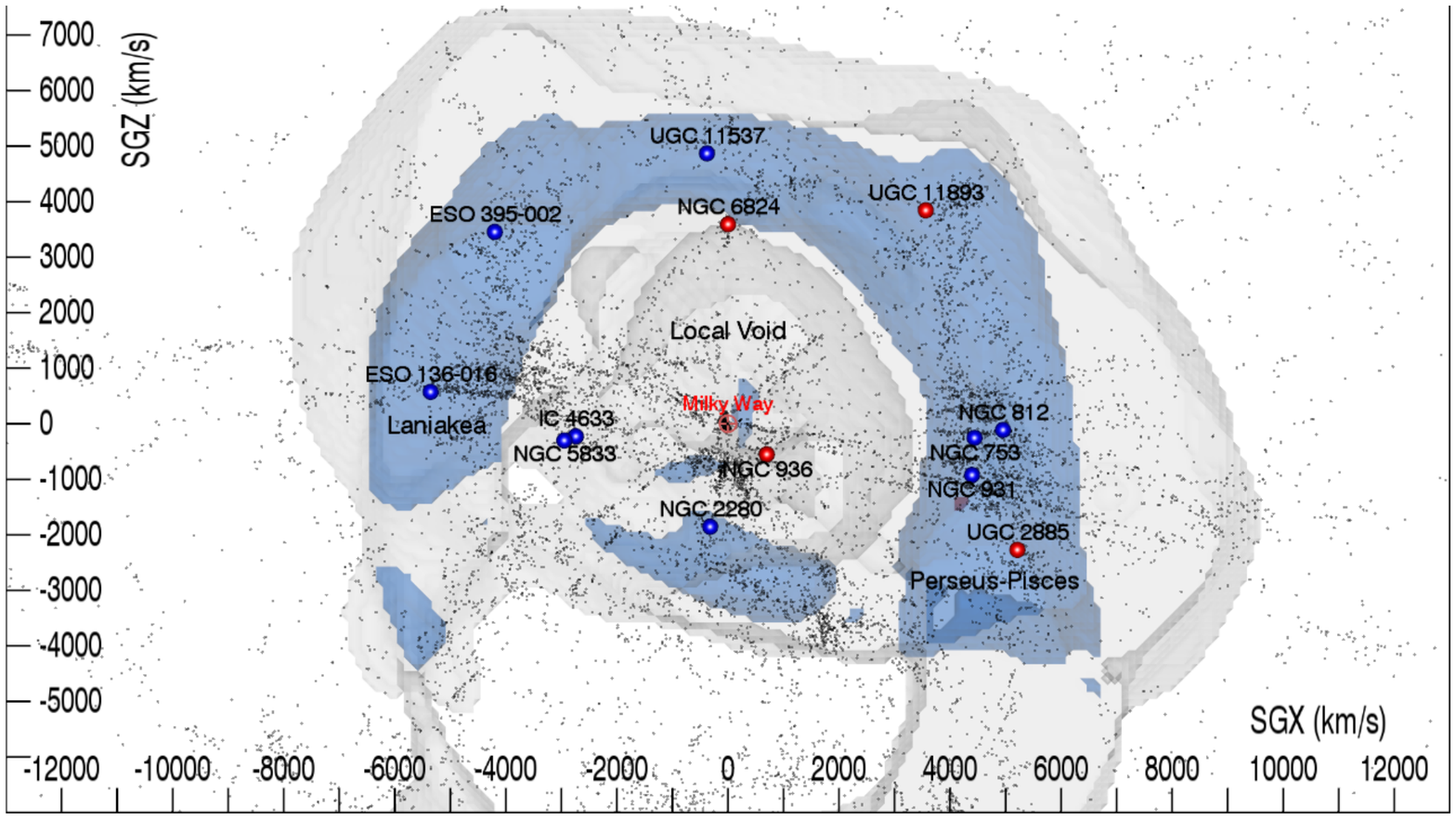}
\vspace{-2cm}
\caption{The environment is studied using the Velocity-Cosmic-Web (V-web). Regions gravitationally collapsing along one direction are sheets, along two orthogonal directions are filaments, along three orthogonal directions are knots. Regions expanding in three directions are voids. Contours in grey, blue and red are showing regions classified respectively as sheets, filaments and knots. The top panel shows a 3D view. The bottom panel is a closer view of a slice of 1000 km/s width on Supergalactic Y axis. The galaxies in 2MASS redshift catalog (2MRS) are plotted with black dots. The cosmography is given with Laniakea on the left and Perseus-Pisces supercluster on the right. The most massive HI galaxies (blue spheres) are located preferentially in filaments with a $\sim$ 100\% higher probability than random galaxies in two different redshift catalogs, V8K or 2MRS. For the fast rotators (red spheres) we cannot confidently identify a preferred environment. }
\label{V-web}
\end{figure}

To pursue this topic further, we explore a novel characterization of environment using the local shear field, as used for the Laniakea discovery \citep{Tully2014Nature}.
A Wiener Filter reconstruction  applied to the Cosmicflows-2 database \citep{2013AJ....146...86T} recovers the underlying 3D velocity field with an effective resolution of a few Megaparsecs \citep{2012ApJ...744...43C}. By taking spatial derivatives of the velocity field, we compute the dimensionless shear tensor. The tensor's eigenvalues ($\lambda_1$, $\lambda_2$, $\lambda_3$)  describe the strength (relative to the Hubble expansion) of compression (positive values) or expansion (negative values) along the eigenvectors of the shear: $\hat{e}_1$, $\hat{e}_2$ , and $\hat{e}_3$. 
At each position in space, we calculate the three eigenvalues of the velocity shear tensor. By ordering these eigenvalues from most positive to most negative, we set thresholds that capture four possibilities. Flows can be inward on all three axes, the condition of a cluster, inward on two axes and outward on the third, the condition of a filament, inward on one axis and outward on two, hence a sheet, or outward on all three axes, hence a void. We define boundaries around contiguous regions with the same shear properties and the contours outline the cosmic web as reconstructed by the V-web algorithm \citep{2012MNRAS.425.2049H}. The statistical uncertainty of the shear eigenvectors and eigenvalues are controlled by means of an ensemble of constrained realizations. The spatial cosmography and reconstructed velocity field of the giant galaxies' environment is shown in Figure~\ref{V-web}.\\ 
We compute the number of galaxies in cells that are classified as knots, filaments, sheets, or nothing (voids or no signal) for two redshift catalogs V8K \citep{Courtois2013} (the most complete redshift catalog within 8,000 km/s)  and 2MRS \citep{2012ApJS..199...26H}.
Some of the studied volume is without signal (for example the zone of avoidance). It is classified as a void in the V-web, but we do not have galaxies there. By confining the analysis to within 8000 km/s, our samples of H{\small I} massive and fast rotators now contain 18 and 17 galaxies, respectively, see Tables 2, 3 and 4.
As seen in Table\ref{Vweb}, our H{\small I} giant galaxies live preferentially in filaments: 10 out of 18 (56\%) while only 21\% of all galaxies within the same volume located today in a filament. Filaments represent 10\% of the volume, contain 21\% of all galaxies, and 56\% of the giants. Adopting a binomial distribution with the probability of a ''true" output (filament) defined to be 0.21, we find that the probability of having 10 or more ''true" outcomes is 0.00025. so the signal is significant at greater than 3$\sigma$ confidence.
In contrast, although our fast rotators also show a hint of a preference for filaments, there the result is only statistically significant at $\sim 1\sigma$ confidence. 

\begin{table*}
\caption{Environmental study using the Cosmic-V-web computed within a sphere of 8,000 km/s radius. Number and percentage (respectively)
of galaxies located in structures dynamically classified as knots, filaments, sheets and voids. The most massive H{\small I} galaxies are preferentially located in filaments,
with a frequency that is a factor of two larger than a random galaxy.}
\begin{tabular}{lrrrr}
\hline
Objects &  in knots & in filaments & in sheets & in void cells \\
\hline
HI massive         &          0  (0\%)  &      10  (56\%)&         4   (22\%)&        4 (22\%)\\
Fast rotators   &      1    (6\%)   &    5    (29\%)  &     8     (47\%)   &   3  (18\%)\\
2MRS galaxies &      586  (2\%)  &    8065  (21\%)&     18236   (47\%)   & 11483 (30\%)\\
V8k galaxies      &   500  (2\%)  &    6408  (21\%)&     14468    (47\%) &   9324 (30\%)\\
Vweb cells        &  2459   (0.2\%)  & 112639  (10\%) &   451487 (42\%)  &    518007  (48\%)\\
\hline
\end{tabular}
\label{Vweb}
\end{table*}

\section{Conclusions}

We investigated the properties of galaxies that potentially challenge the current paradigm of galaxy evolution. Albeit rare, these massive disk galaxies run counter to the idea that mass alone determines a galaxy's star formation rate or its morphology. Of course there are other previously known reasons to disfavor such an extreme view, but well-specified, carefully defined samples have the potential to highlight these discrepancies. We selected  the 100 most H{\small I} massive galaxies and the 100 most rapidly rotating disk galaxies from H{\small I} catalogs available for public use at the Extragalactic Distance Database that contains thousands of coherently measured H{\small I} profiles.

To compare with the overall disk population,  we use the baryonic Tully-Fisher (BTF) relation as a benchmark. That relation holds broadly because a nearly constant fraction of the available baryons in each galaxy's halo settle into the disk \citep{zbtf}. The classic Tully-Fisher (TF) relation then holds because for most galaxies a fixed fraction of those settled, or ``condensed", baryons turn into stars. Deviations from the BTF therefore indicate a variation in the fraction of condensed baryons, $f_C$, which could signal interesting evolutionary deviations from the typical galaxy. Deviations from the TF relation could similarly signal deviations $f_C$, or they could signal deviations in the star formation efficiency. 

Using the H{\small I} data in combination with available 3.6$\mu$m photometry from {\sl Spitzer} observations, we place our two samples on the BTF. We find both samples deviate systematically from the existing BTF.  The rapid rotators tend to lie below the BTF. We discuss a number of possible reasons for this deviation, including a suggestion that the scaling relation is not linear at the massive end. Such behavior is also seen in massive early-type galaxies \citep{bernardi}. The systematic deviation might also hint at bottom heavy IMFs in more massive galaxies, again as seen in massive early type galaxies \citep{vandokkum,cappellari}. 

The H{\small I} massive galaxies lie systematically above the BTF. If these deviations are interpreted as variations in $f_C$, then we conclude that there are galaxies in which $f_C$ approaches 1, or in other words the entire allotment of baryons for that halo have found their way to the disk. In spite of this, or whatever other process may be driving the deviation from the BTF, the star formation efficiency, when quantified as the fraction of disk baryons that are in stars, is the same in these systems as in more typical galaxies. We conclude that the star formation efficiency is driven by internal self-regulation, while $f_C$ can be altered by external factors. 

Although mass is clearly not the driver of the morphological or star formation history differences that exist between these galaxies and other similarly massive galaxies, we have not identified the driver. Our galaxies avoid the densest environments, but so do other galaxies that are not similarly extreme. When looking in detail for environmental differences, we found 
that the most massive HI detected galaxies are located preferentially in filaments.
We present the first evidence of an environmental effect on galaxy evolution using a dynamical definition of a filament.

\section{Acknowledgments}

This research has made use of the NASA/IPAC Extragalactic
Database (NED) which is operated by the Jet Propulsion
Laboratory, California Institute of Technology, under
contract with the National Aeronautics and Space
Administration. We acknowledge the usage of the HyperLeda
database (http://leda.univ-lyon1.fr). HC and
JS acknowledge support from the Lyon Institute of Origins
under grant ANR-10-LABX-66 and from CNRS under
PICS-06233. DZ acknowledges support from a NASA ADAP award (NNX12AE27G) and
an NSF AAG grant (AST-1311326).
We thank Yehuda Hoffman from the Hebrew University in Jerusalem for providing a Cosmic-V-web computation of the cosmicflows-2 database.

\clearpage

\begin{table*}
\caption{The 20 largest H{\small I}-mass galaxies}
\begin{tabular}{ccccccccccc}
\hline
PGC & Name & J2000 coordinates & Vhel  & $W_{mx}$ & error on $W_{mx}$ & inclination & $V_{max}$& $M_{H{\small I}}/M_{\odot}$ & morph. type  & $log(d_{25})$   \\
         &             &                                & $km\,s^{-1}$& $km\,s^{-1}$& $km\,s^{-1}$ & degrees & $km\,s^{-1}$ & 10$^{10}$ solar mass units& & in log of 0.1 arcmin\\
\hline
  17625 &         NGC1961 &     J054204.8+692243 &   3935 & 658 &   8 &  47.0 &   449.86 & 4.46 &    4.2 &   1.65 \\
   9560 &         NGC0958 &     J023042.8-025621 &   5739 & 570 &  10 &  78.1 &   291.26 & 3.77 &    4.9 &   1.40 \\
  14030 &        UGC02885 &     J035302.4+353522 &   5804 & 556 &   8 &  62.4 &   313.70 & 3.35 &    5.2 &   1.63 \\
   8066 &         NGC0812 &     J020651.6+443418 &   5161 & 435 &  16 &  71.9 &   228.82 & 3.24 &    5.9 &   1.44 \\
  64458 &        UGC11537 &     J201838.1-000902 &   5418 & 499 &   8 &  69.7 &   266.02 & 3.09 &    5.3 &   1.18 \\
  34836 &         NGC3646 &     J112143.1+201011 &   4246 & 508 &  16 &  63.9 &   282.84 & 2.96 &    4.6 &   1.49 \\
  52328 &         NGC5720 &     J143833.3+504855 &   7784 & 409 &  17 &  52.1 &   259.16 & 2.80 &    3.0 &   1.28 \\
  56891 &      ESO136-016 &     J160349.3-605840 &   5426 & 521 &  12 &  90.0 &   260.50 & 2.76 &    5.2 &   1.52 \\
  24830 &        UGC04625 &     J085017.8+032951 &   8476 & 422 &  18 &  90.0 &   211.00 & 2.74 &    6.1 &   1.23 \\
  62178 &         NGC6674 &     J183833.9+252230 &   3429 & 431 &  16 &  62.9 &   242.08 & 2.63 &    3.0 &   1.60 \\
  16537 &        UGC03218 &     J050043.7+621439 &   5228 & 495 &  15 &  56.4 &   297.15 & 2.62 &    3.1 &   1.24 \\
  59884 &          IC4633 &     J171346.9-773210 &   2945 & 318 &  13 &  62.4 &   179.42 & 2.57 &    6.0 &   1.43 \\
  61791 &      ESO395-002 &     J182226.4-354040 &   5607 & 538 &  13 &  90.0 &   269.00 & 2.56 &    3.4 &   1.32 \\
  19531 &         NGC2280 &     J064449.1-273819 &   1897 & 384 &   7 &  66.2 &   209.84 & 2.49 &    5.9 &   1.81 \\
   7387 &         NGC0753 &     J015742.2+355457 &   4902 & 314 &   8 &  52.5 &   197.89 & 2.41 &    4.9 &   1.14 \\
  52361 &        UGC09437 &     J143910.6+184247 &  14401 & 404 &  16 &  65.6 &   221.81 & 2.38 &    3.3 &   0.90 \\
  54250 &         NGC5833 &     J151153.6-725134 &   3031 & 407 &  14 &  76.7 &   209.11 & 2.36 &    4.2 &   1.49 \\
   9399 &         NGC0931 &     J022814.5+311840 &   4999 & 422 &   9 &  81.3 &   213.45 & 2.30 &    3.6 &   1.39 \\
  18089 &        UGC03375 &     J055525.3+515438 &   5791 & 490 &  11 &  64.2 &   272.13 & 2.29 &    5.2 &   1.18 \\
  28196 &         NGC2998 &     J094843.6+440453 &   4772 & 374 &   7 &  61.8 &   212.18 & 2.27 &    5.2 &   1.39 \\	   
\hline
\end{tabular}

\label{tb1}
\end{table*}

\clearpage
\begin{table*}
\caption{The 20 fastest rotators}
\begin{tabular}{ccccccccccc}
\hline
PGC & Name & J2000 coord. & Vhel & $W_{mx}$ & error on $W_{mx}$ & inclination & $V_{max}$ & $M_{H{\small I}}/M_{\odot}$ & morph. type  &  $log(d_{25}) $ \\
         &              &                        &  $km\,s^{-1}$ &  $km\,s^{-1}$ &  $km\,s^{-1}$ & degrees & $km\,s^{-1}$ & 10$^{10}$ solar mass units &  & log of 0.1 arcmin\\
\hline
  17625 &         NGC1961 &     J054204.8+692243 &   3935 & 658 &   8 &  47.0 &   449.86 & 4.46    &    4.2  &  1.65 \\
  42407 &         NGC4594 &     J123959.4-113723 &   1095 & 733 &  17 &  59.4 &   425.80 & 0.0430  &    1.1  &  1.93 \\
  63575 &         NGC6824 &     J194340.7+560634 &   3549 & 544 &  12 &  45.8 &   379.41 & 0.148   &    2.3  & 1.28 \\
   9359 &         NGC0936 &     J022737.4-010921 &   1436 & 545 &  11 &  50.5 &   353.15 & 0.0907  &   -1.2  &  1.65 \\
 165398 &       PGC165398 &     J043157.1+592547 &   4626 & 489 &  20 &  45.5 &   342.80 & 0.440   &    0.0  &  0.94 \\
  58596 &         NGC6195 &     J163632.6+390141 &   9021 & 517 &  18 &  50.2 &   336.47 & 0.221  &    3.1  &  1.16 \\
  65375 &         NGC6962 &     J204719.1+001915 &   4222 & 475 &  13 &  45.0 &   335.88 & 1.12    &    1.7  &  1.43 \\
  24685 &      ESO563-021 &     J084717.0-200208 &   4582 & 664 &   7 &  90.0 &   332.00 & 2.17    &    4.3  &  1.48 \\
  26512 &         NGC2841 &     J092202.5+505837 &    633 & 592 &  12 &  65.2 &   326.07 & 0.236   &    3.0  &  1.84 \\
  25161 &         NGC2713 &     J085720.5+025517 &   3916 & 629 &  18 &  77.7 &   321.89 & 0.503   &    2.5  &   1.53 \\
  36706 &         NGC3884 &     J114612.2+202330 &   6947 & 462 &  20 &  45.9 &   321.67 & 0.660   &    0.6  &   1.28 \\
  52665 &         NGC5746 &     J144456.0+015717 &   1723 & 634 &  11 &  90.0 &   317.00 & 0.456   &    3.0  &   1.86 \\
  57173 &        UGC10205 &     J160640.2+300556 &   6562 & 541 &  20 &  59.0 &   315.58 & 0.725   &    1.0  &   1.16 \\
  14030 &        UGC02885 &     J035302.4+353522 &   5804 & 556 &   8 &  62.4 &   313.70 & 3.35    &    5.2  &   1.63 \\
  37617 &         NGC3992 &     J115735.9+532228 &   1048 & 459 &   8 &  47.4 &   311.78 & 0.218   &    4.0  &   1.91 \\
  72233 &        UGC12755 &     J234349.7+282021 &   8794 & 503 &  19 &  54.7 &   308.16 & 0.648   &    3.1  &   1.07 \\
  67966 &        UGC11893 &     J220406.7+355618 &   5589 & 606 &  20 &  80.2 &   307.48 & 0.924   &    6.3  &   0.81 \\
  66880 &        UGC11758 &     J213057.6+135910 &   8635 & 612 &  19 &  90.0 &   306.00 & 0.942   &    4.2  &   1.12 \\
  45947 &         NGC5032 &     J131327.0+274808 &   6413 & 524 &  19 &  59.5 &   304.08 & 0.212   &    3.0  &  1.29 \\
  63286 &        UGC11455 &     J192956.3+720646 &   5392 & 608 &  15 &  90.0 &   304.00 & 1.29    &    5.8  &  1.39 \\
 \hline
\end{tabular}

\label{tb2}
\end{table*}

\clearpage

\begin{table*}		   
\caption{HI-21cm spectrum and Spitzer $3.6 \mu$ images of the top 20 most heavy HI-mass galaxies, in decreasing mass order.}
\begin{tabular}{cccc}
            
\hline
PGC - Name & 
 HI mass ($log(M_{HI}$) in solar units and V$_{hel}$ & 
 PGC - Name & 
  HI mass ($log(M_{HI}$) in  solar units and V$_{hel}$ \\
Telescope / HI profile &
  Spitzer 3.6 $\mu$ IRAC 400X400 pixels &  
   Telescope / HI profile & 
    Spitzer 3.6 $\mu$ IRAC 400X400 pixels \\
  \hline  
    \hline  
PGC17625	& 	4.46 / 3935~$km\,s^{-1}$  &    PGC9560 & 	3.77 /  5739~$km\,s^{-1}$ \\
 \includegraphics[width=4cm]{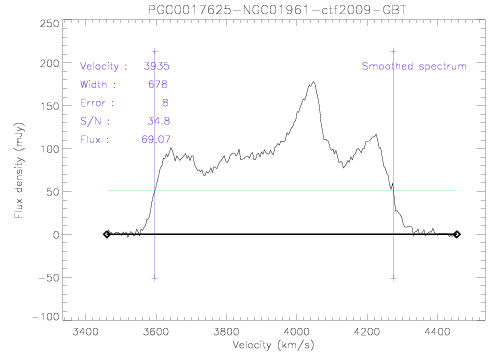}&
 \includegraphics[width=3cm]{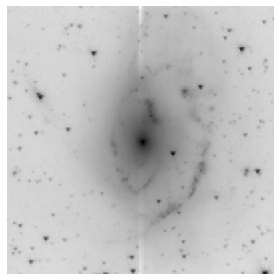}&
  \includegraphics[width=4cm]{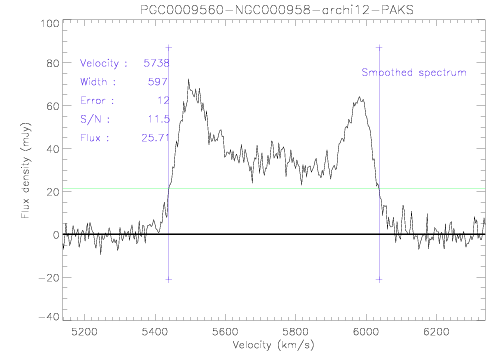}&
 \includegraphics[width=3cm]{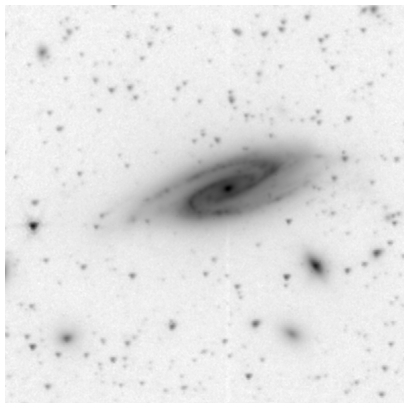}\\
GBT  & & Parkes  64m\\
   \hline  
   PGC14030	& 	  3.35 / 5804~$km\,s^{-1}$  &    
PGC8066 & 	  3.24 / 5161~$km\,s^{-1}$ \\
 \includegraphics[width=4cm]{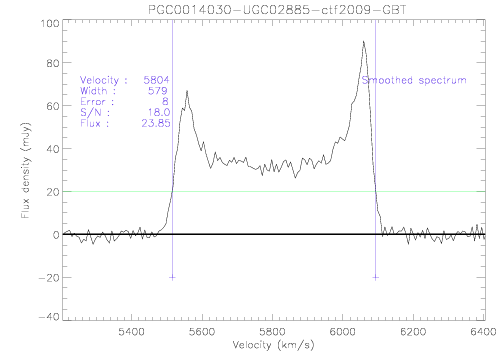}&
 \includegraphics[width=3cm]{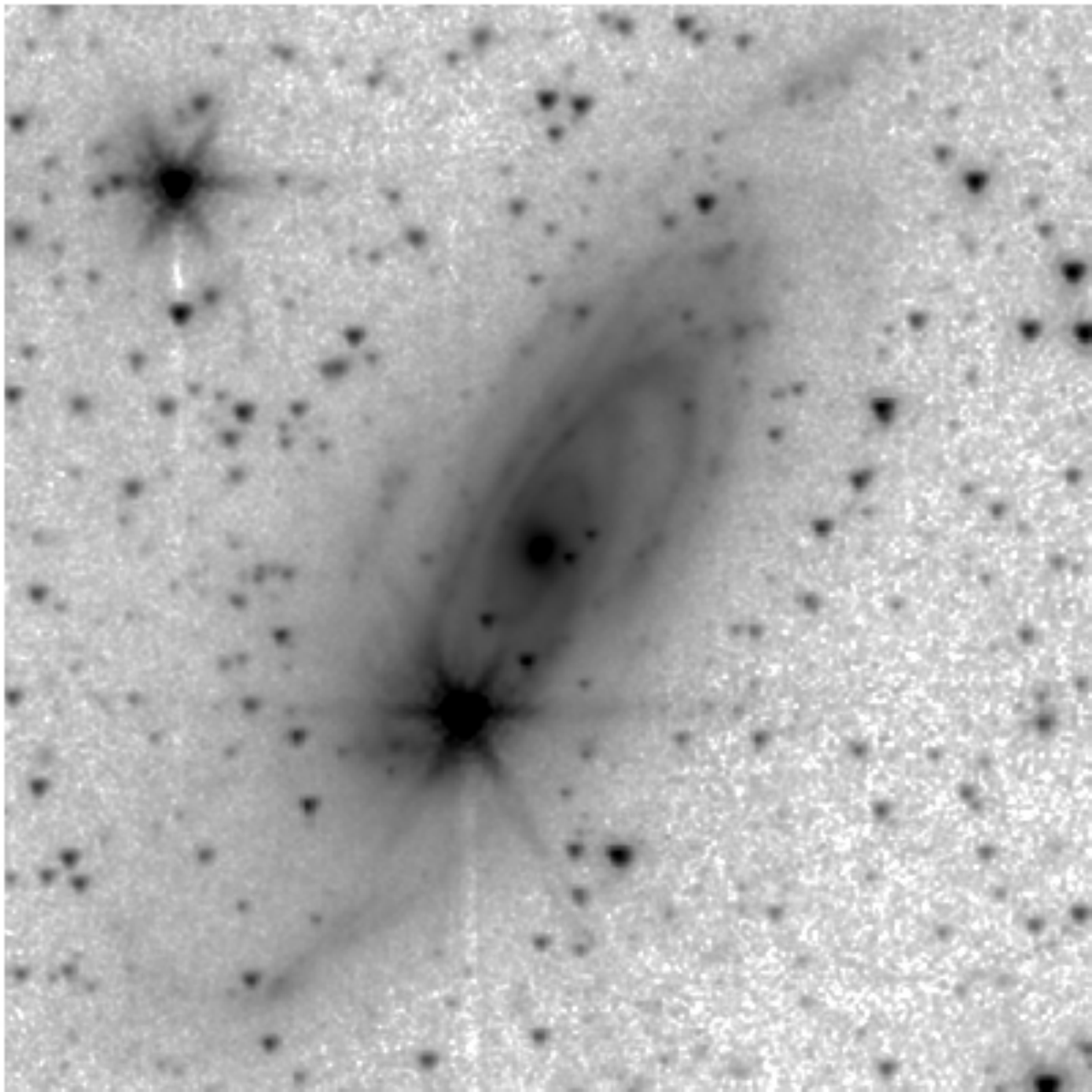}&
  \includegraphics[width=4cm]{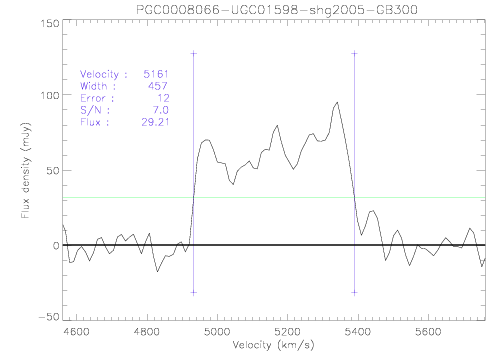}&
 \includegraphics[width=3cm]{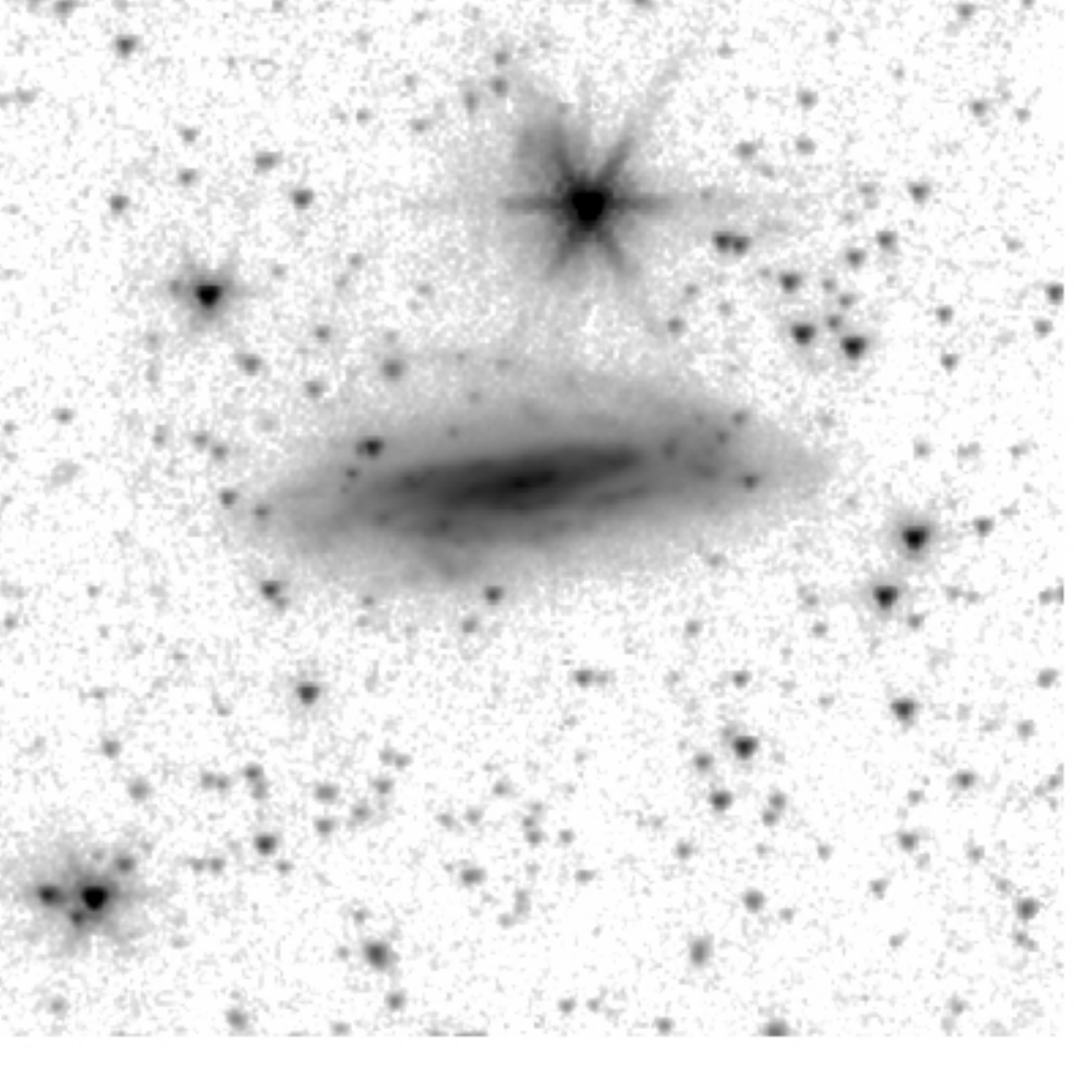}\\
 GBT && GB300 \\
   \hline  
PGC64458	& 	 3.09 / 5418 ~$km\,s^{-1}$  &    
PGC34836 & 	 2.96 / 4246 ~$km\,s^{-1}$ \\
 \includegraphics[width=4cm]{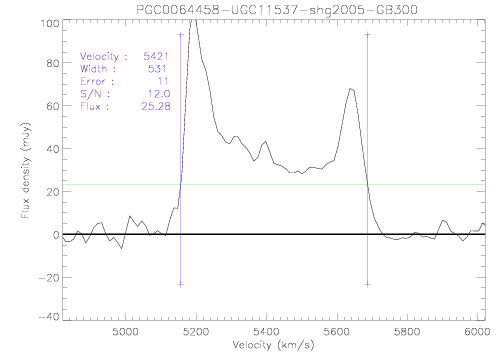}&
 \includegraphics[width=3cm]{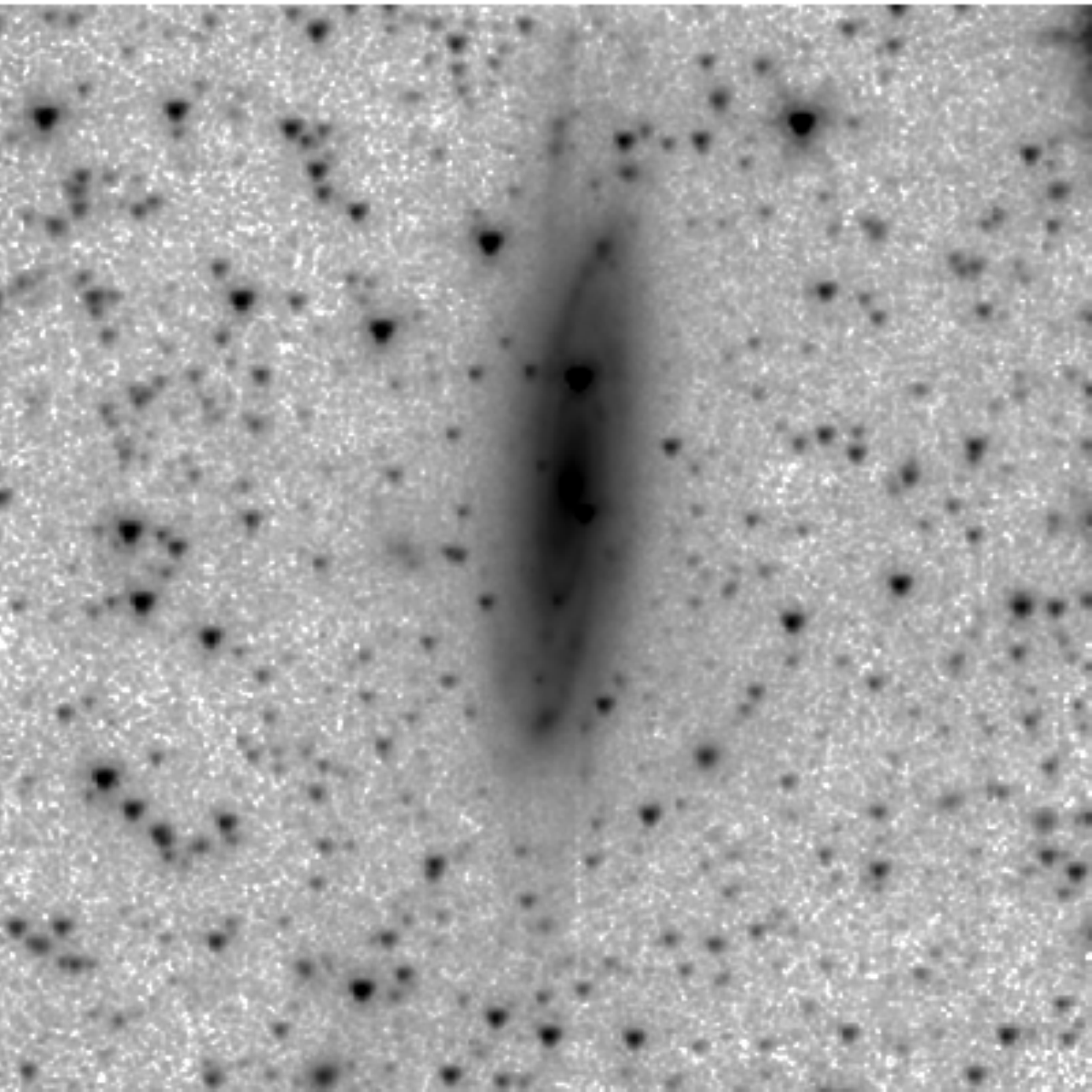}&
  \includegraphics[width=4cm]{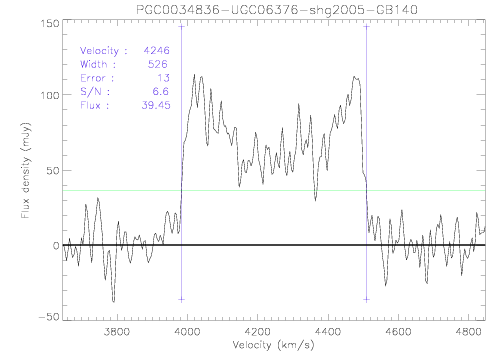}&
 \includegraphics[width=3cm]{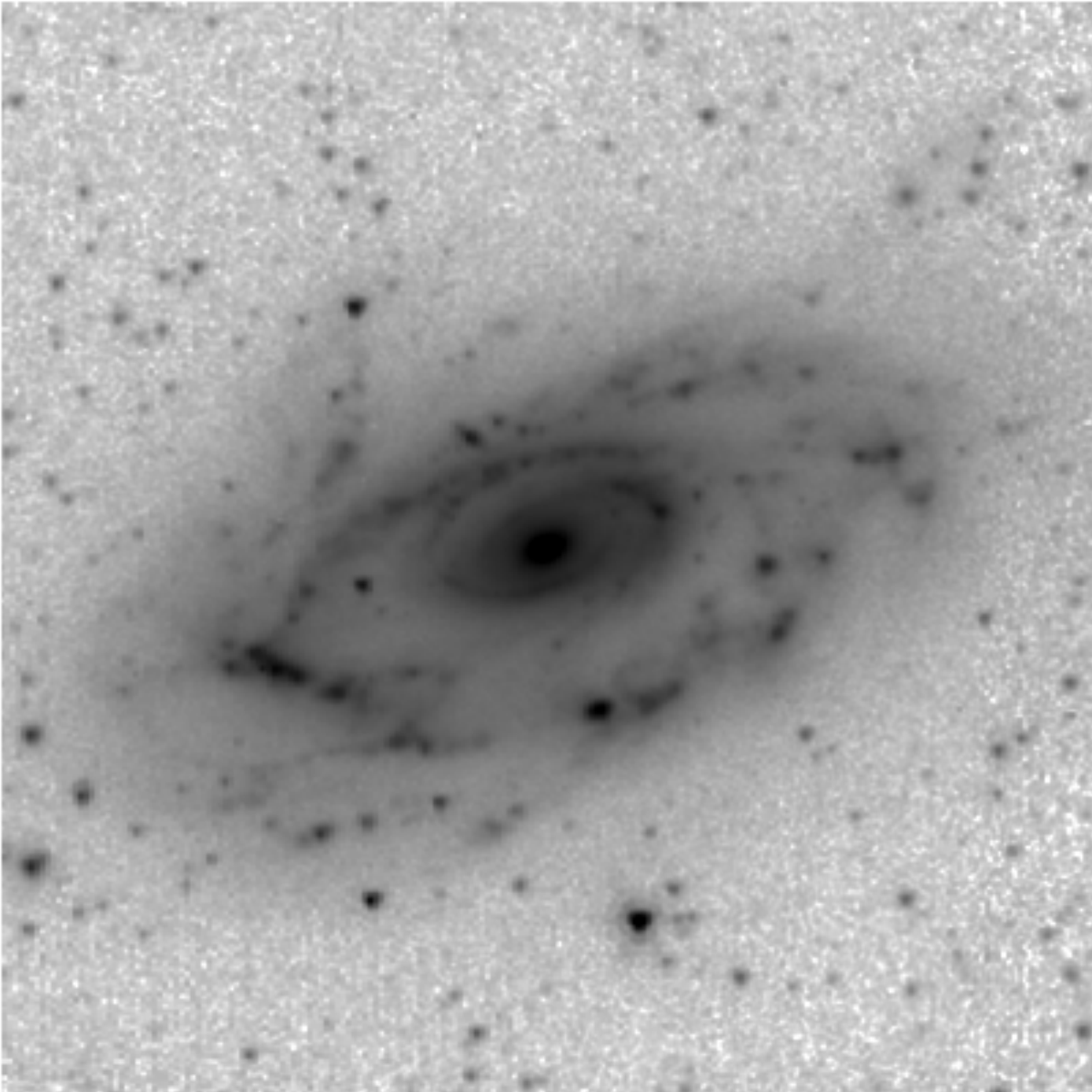}\\
 GB300 && GB140 \\
   \hline  

PGC52328  	& 	 2.80 / 7784 ~$km\,s^{-1}$  &    
PGC56891 & 	  2.76 / 5426~$km\,s^{-1}$ \\
 \includegraphics[width=4cm]{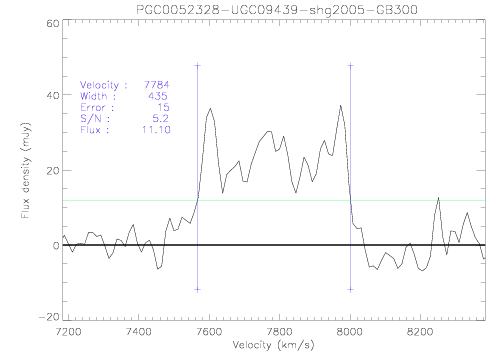}&
 \includegraphics[width=3cm]{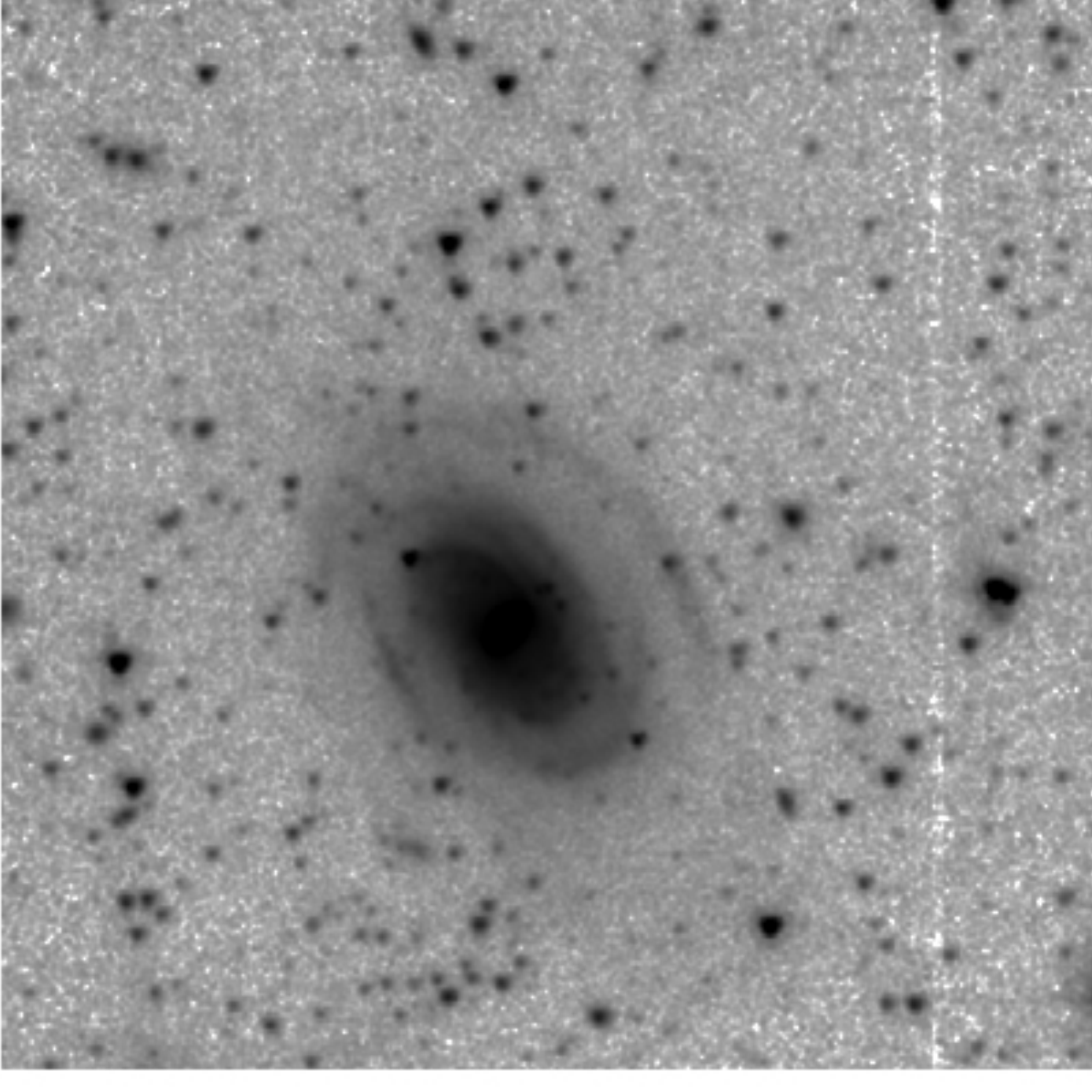}&
  \includegraphics[width=4cm]{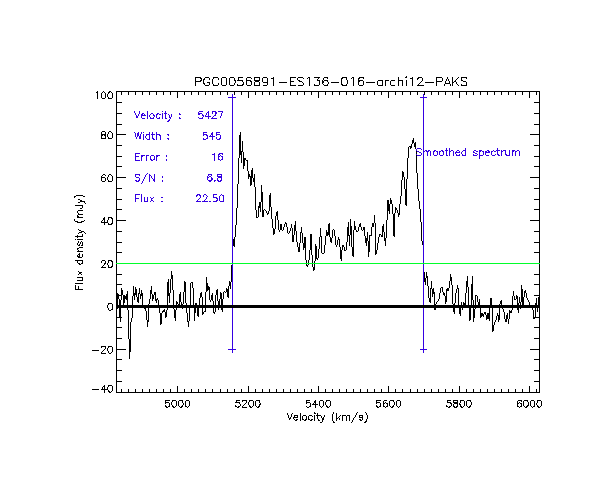}&
 \includegraphics[width=3cm]{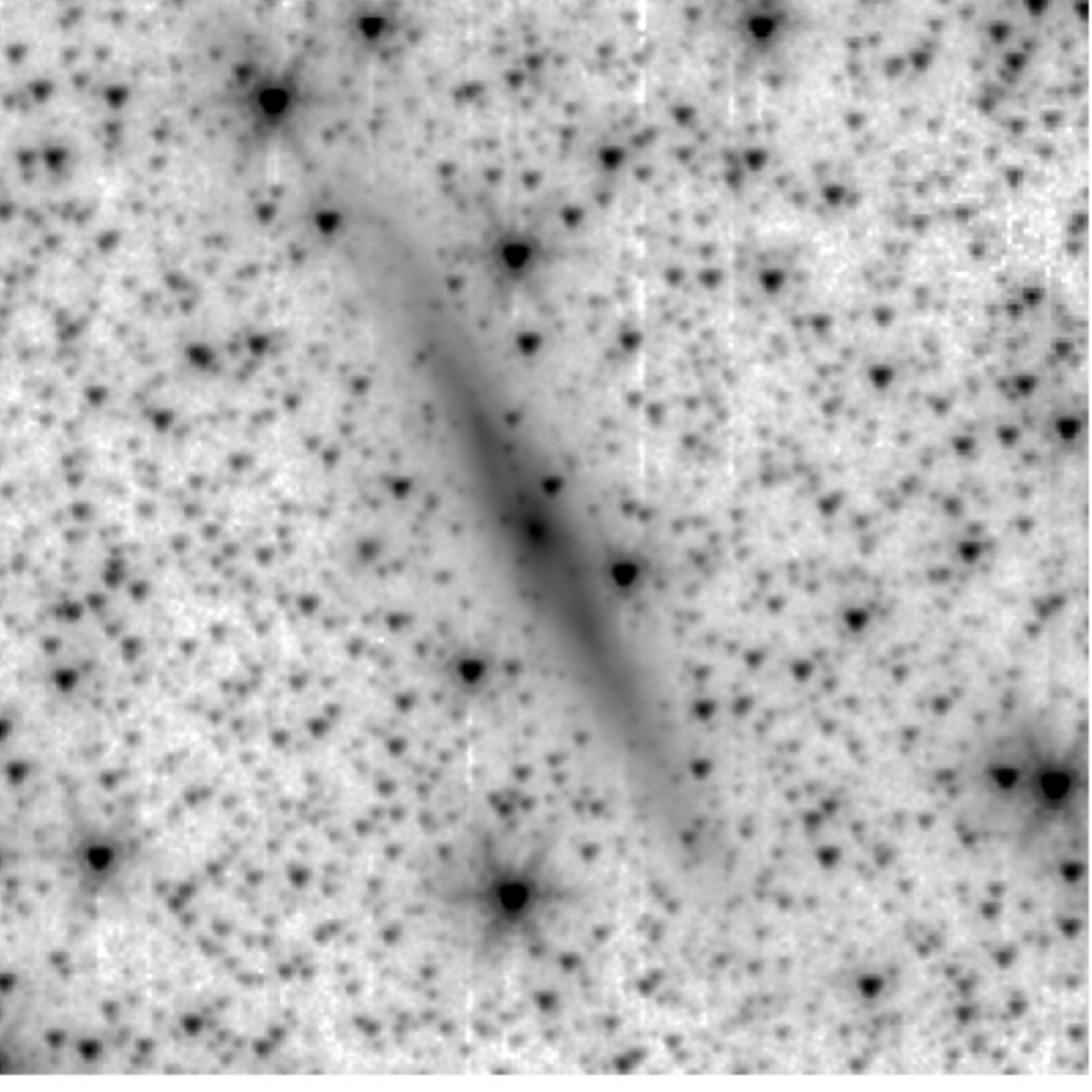}\\
 GB300 && Parkes 64 m\\
 \hline
PGC24830	& 	 2.74 / 8476 ~$km\,s^{-1}$  &    
PGC62178   & 	  2.63 / 3429~$km\,s^{-1}$ \\
 \includegraphics[width=4cm]{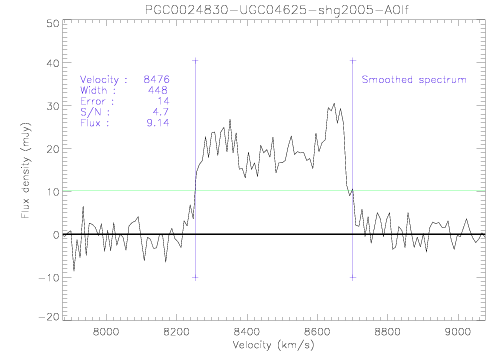}&
 \includegraphics[width=3cm]{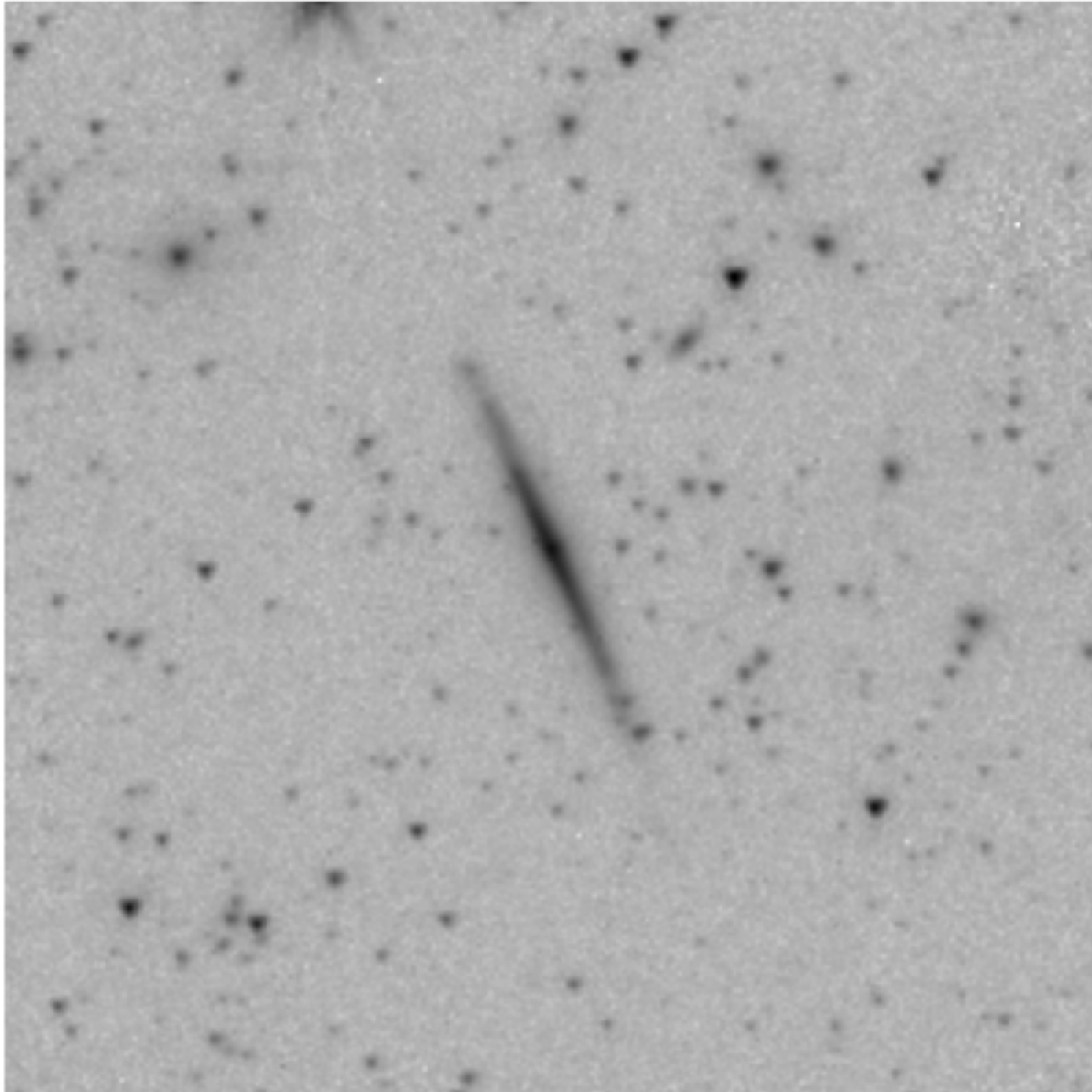}&
  \includegraphics[width=4cm]{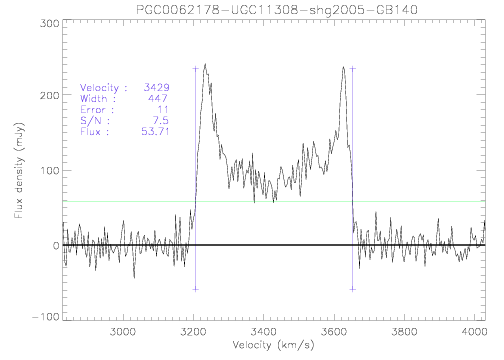}&
 \includegraphics[width=3cm]{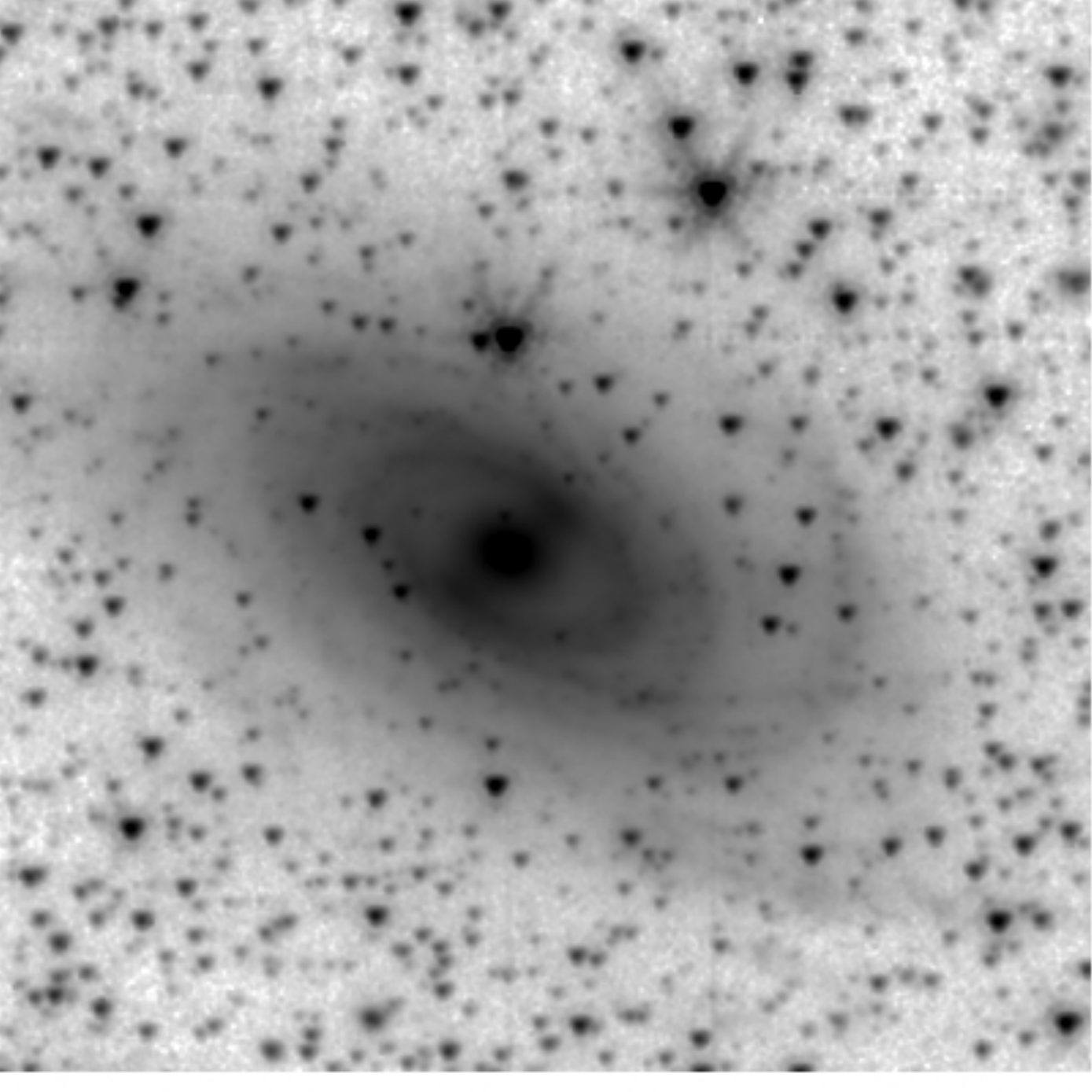}\\
  Arecibo && GB140 \\
  \hline  
\end{tabular}
\end{table*}
\clearpage
\newpage
\begin{table*}		   
\caption{Table4 .. continued ...HI-21cm spectrum and Spitzer $3.6 \mu$ images of the top 20 most heavy HI-mass galaxies, in decreasing mass order.}
\begin{tabular}{cccc}
            \hline
PGC - Name & HI mass ($log(M_{HI}$) in  solar units and V$_{hel}$   - & PGC - Name & HI mass ($log(M_{HI}$) in solar units and V$_{hel}$\\
  Telescope / HI profile & Spitzer 3.6 $\mu$ IRAC 400X400 pixels &  Telescope / HI profile & Spitzer 3.6 $\mu$ IRAC 400X400 pixels \\
    \hline  
PGC16537	& 	 2.62 / 5228 ~$km\,s^{-1}$  &    
PGC59884  & 	  2.57 / 2945~$km\,s^{-1}$ \\
 \includegraphics[width=4cm]{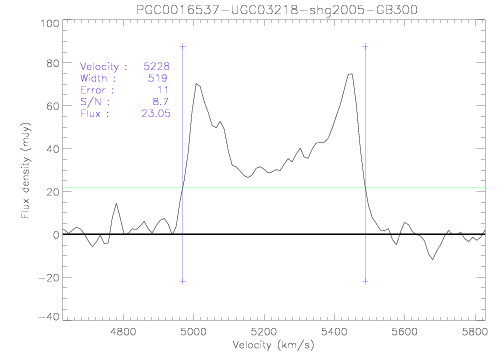}&
 \includegraphics[width=3cm]{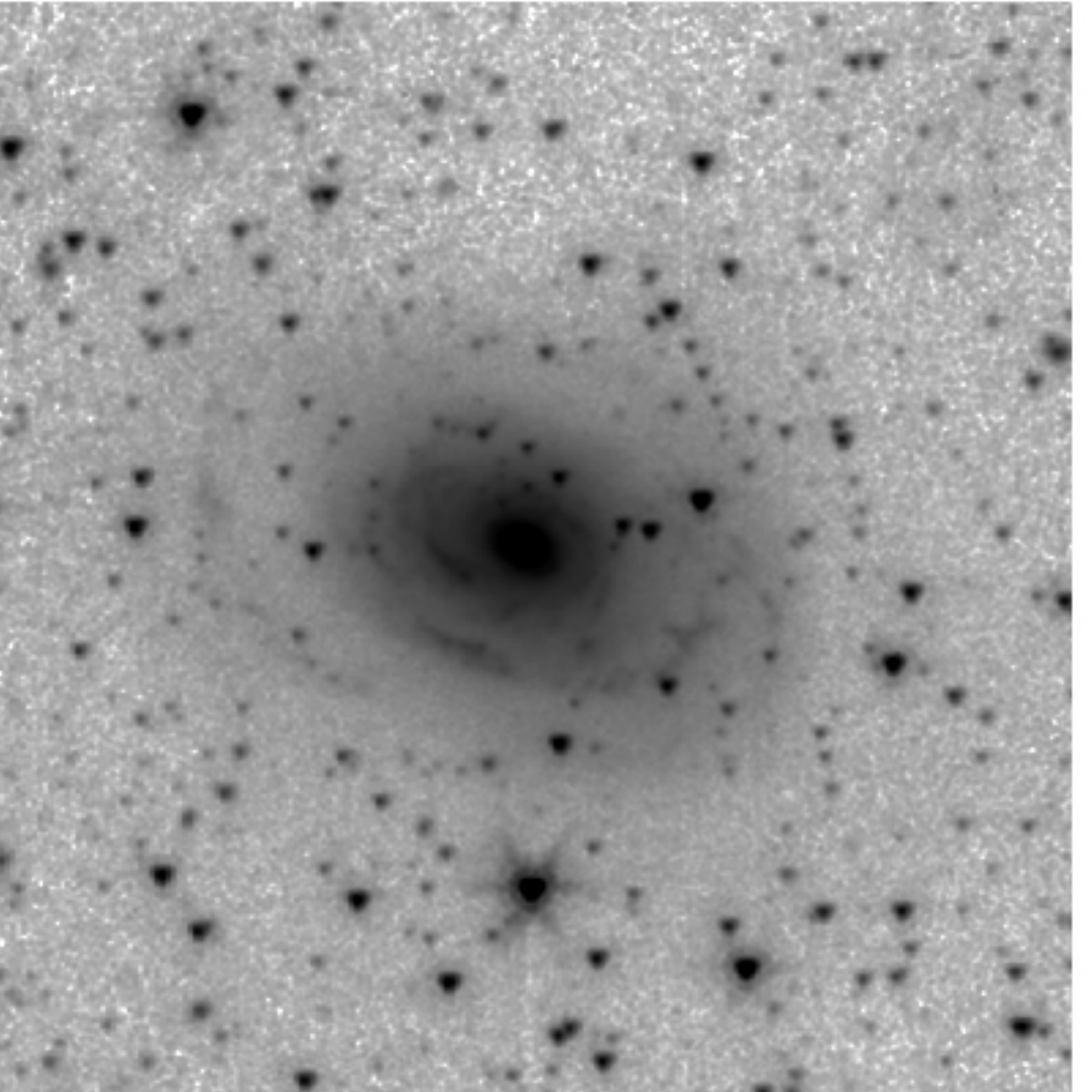}&
  \includegraphics[width=4cm]{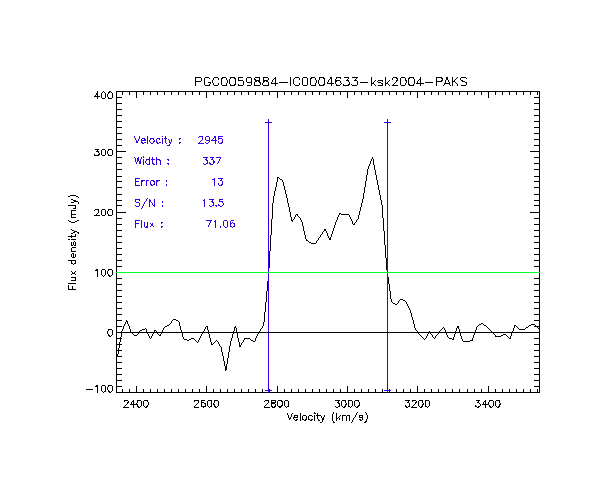}&
 \includegraphics[width=3cm]{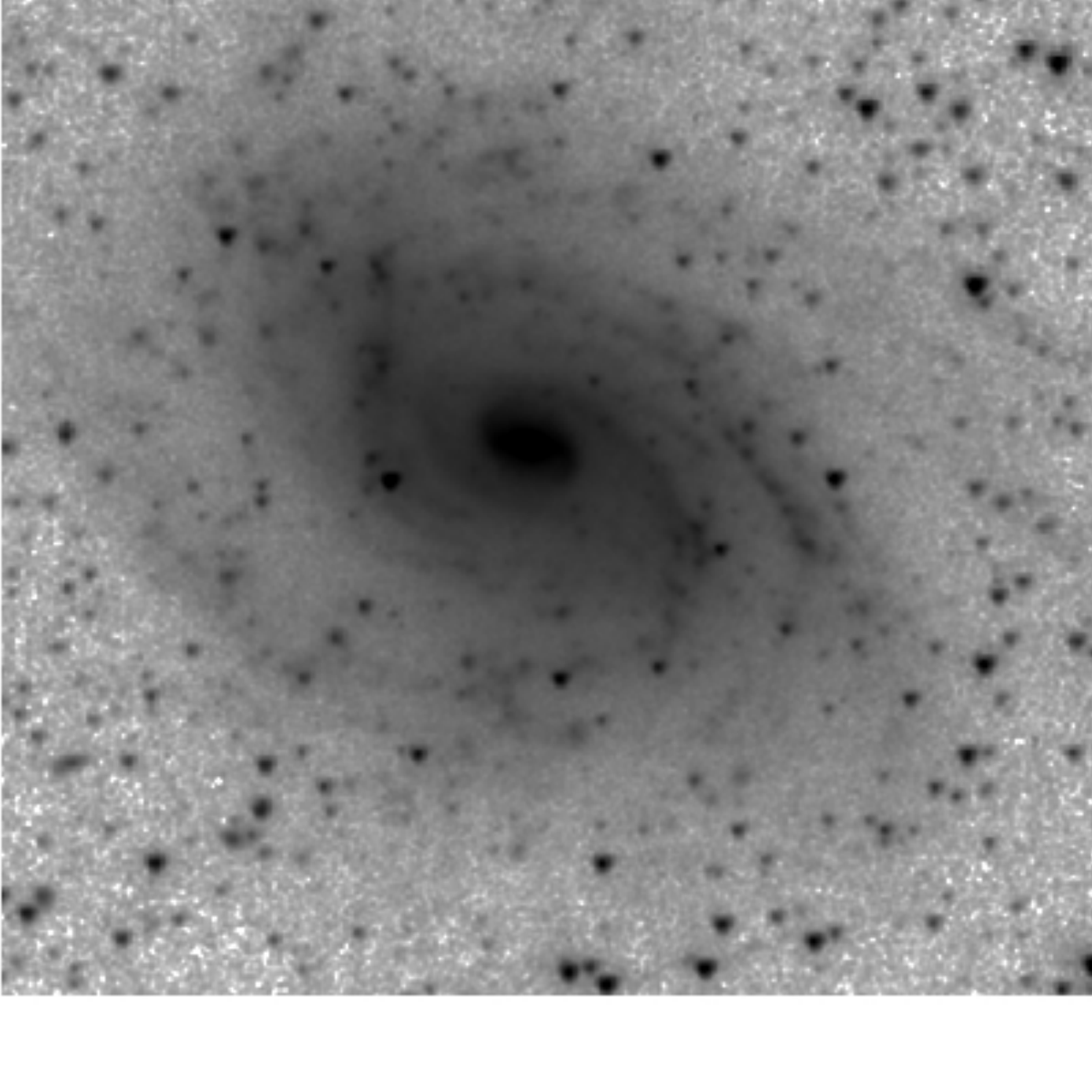}\\
GB300   && Parkes 64m \\
   \hline  
     \hline  
PGC61791	& 	  2.56 / 5607~$km\,s^{-1}$  &    
PGC19531   & 	  2.49 / 1897~$km\,s^{-1}$ \\
 \includegraphics[width=4cm]{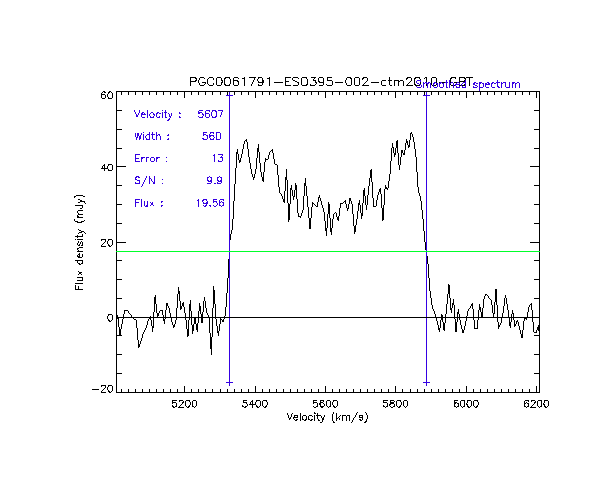}&
 \includegraphics[width=3cm]{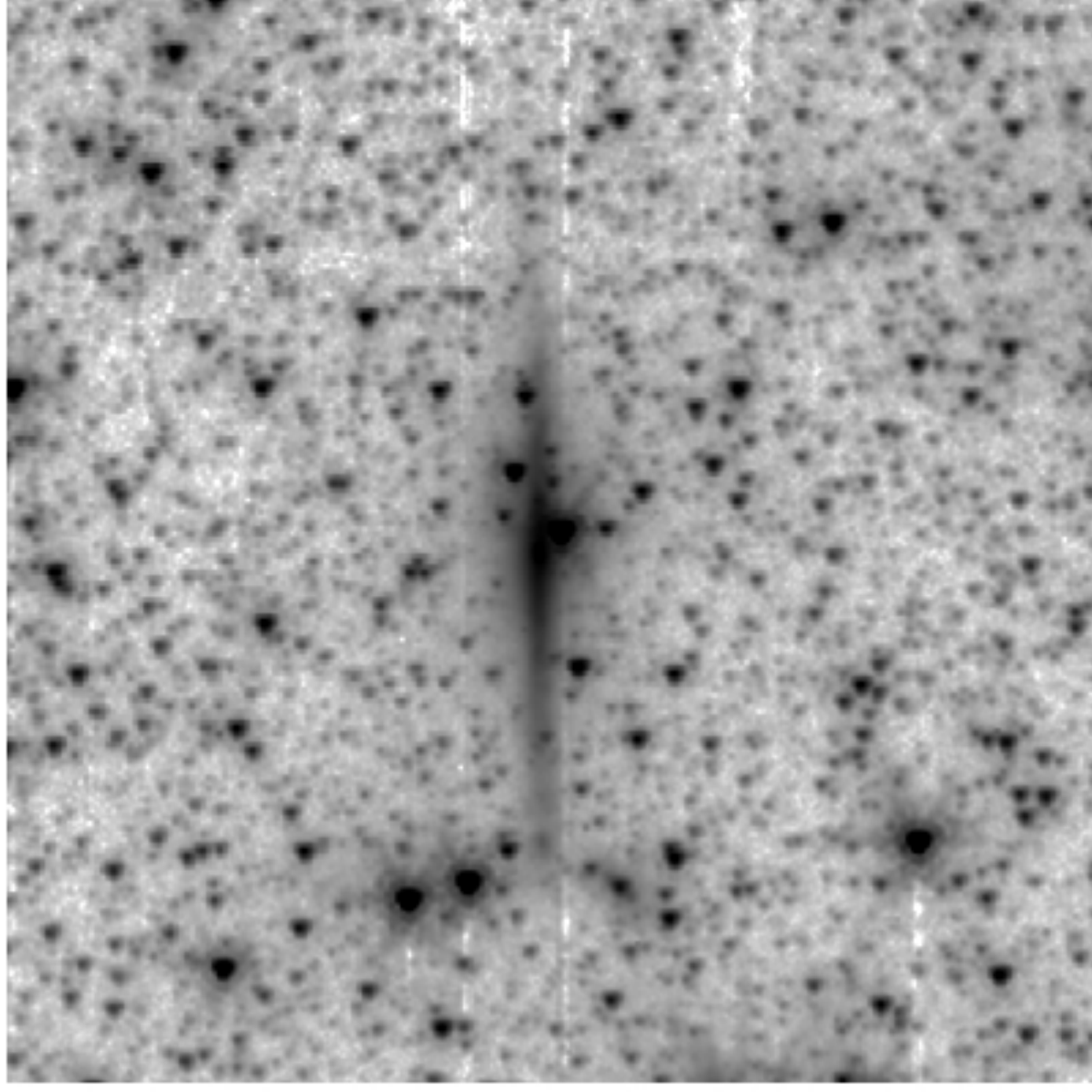}&
  \includegraphics[width=4cm]{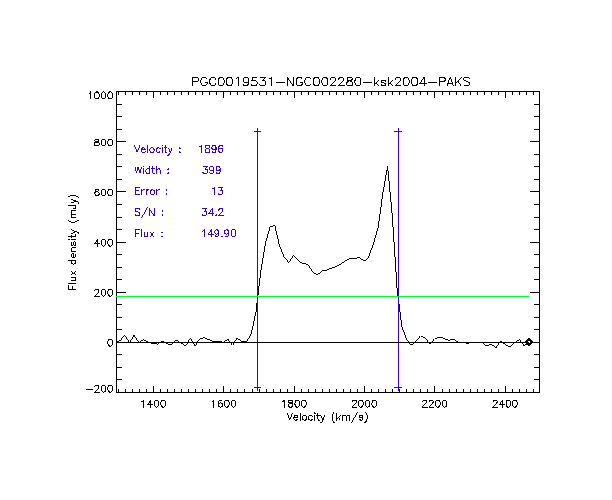}&
 \includegraphics[width=3cm]{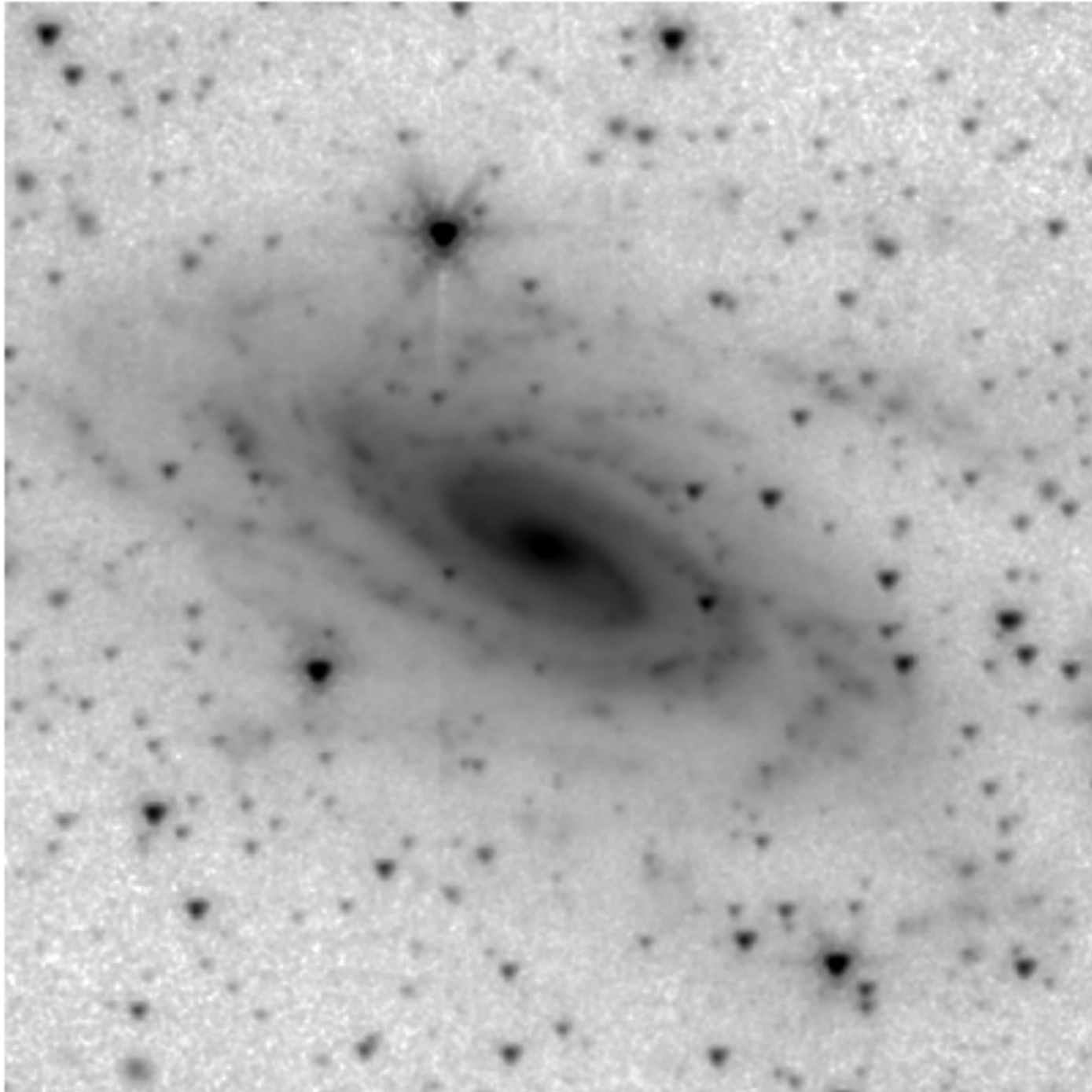}\\
 GBT  && Parkes 64m\\
   \hline  
     \hline  
PGC7387  	& 	 2.41 / 4902 ~$km\,s^{-1}$  &    
PGC52361  & 	  2.38 / 14401~$km\,s^{-1}$ \\
 \includegraphics[width=4cm]{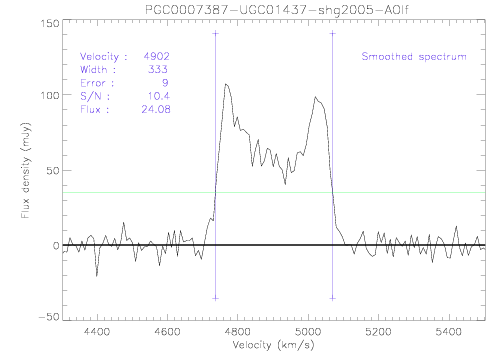}&
 \includegraphics[width=3cm]{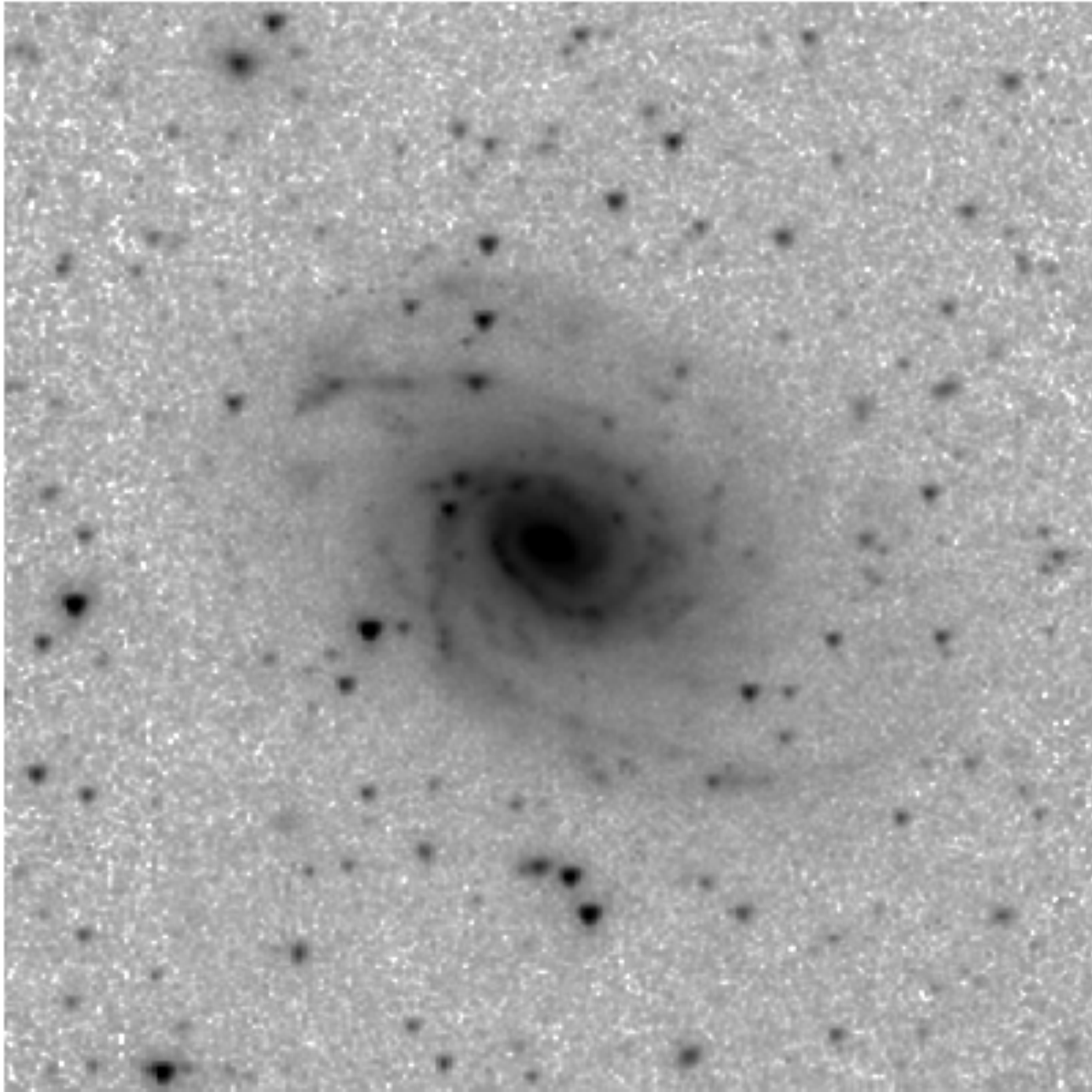}&
  \includegraphics[width=4cm]{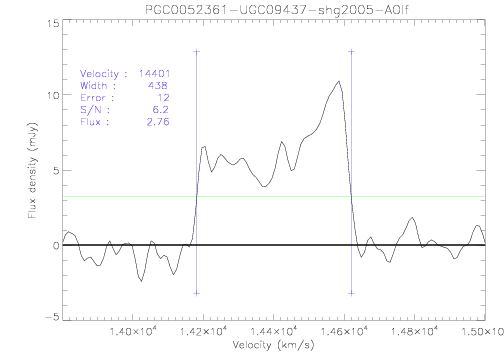}&
 \includegraphics[width=3cm]{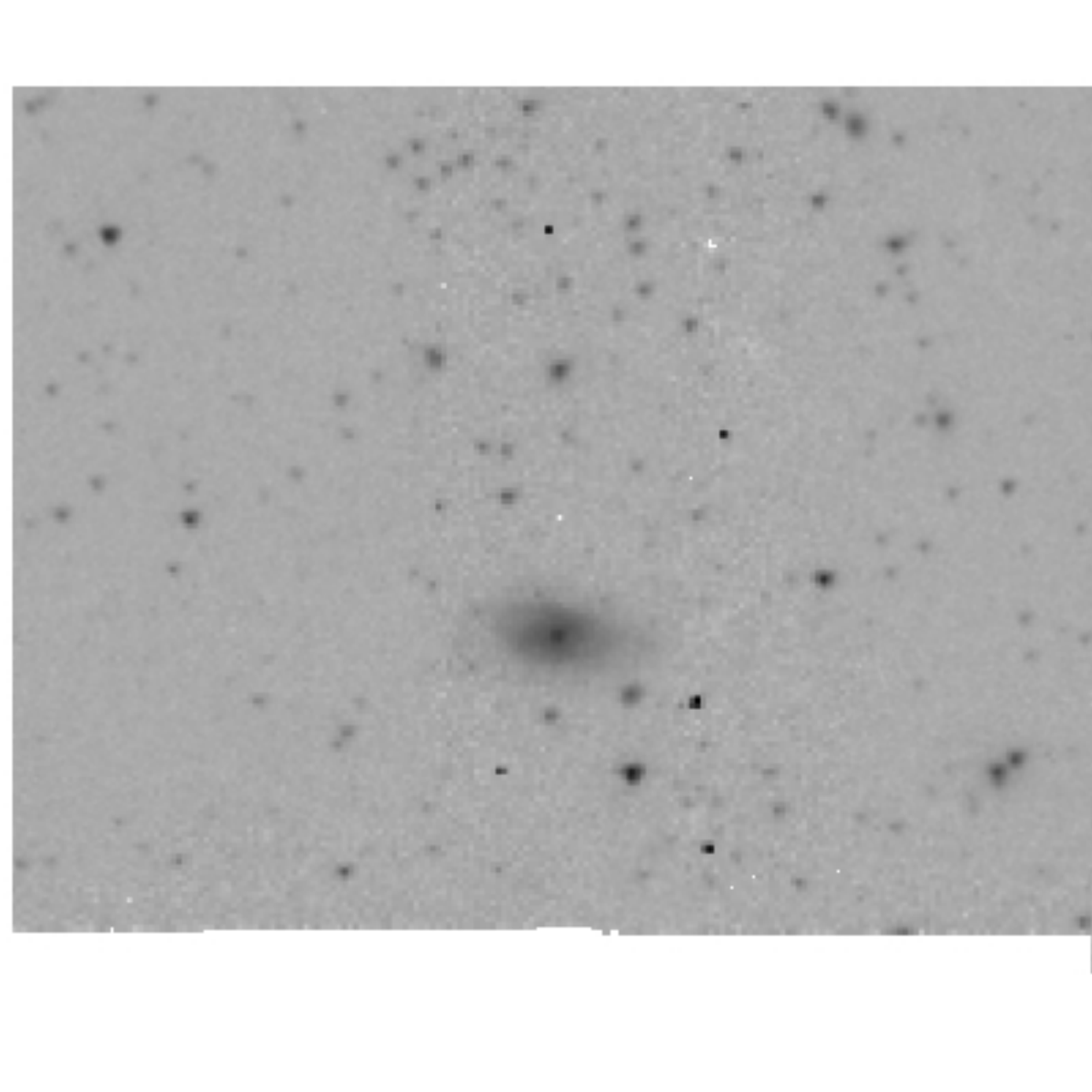}\\
Arecibo  && Arecibo\\
   \hline  
     \hline  
PGC54250	& 	  2.36 / 3031~$km\,s^{-1}$  &    
PGC9399  & 	  2.30 / 4999~$km\,s^{-1}$ \\
 \includegraphics[width=4cm]{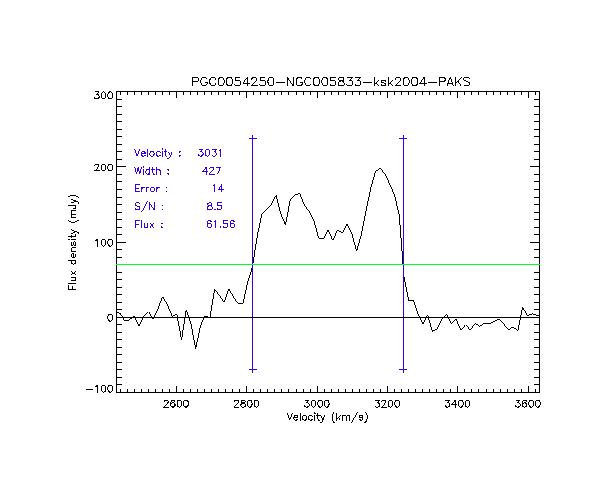}&
 \includegraphics[width=3cm]{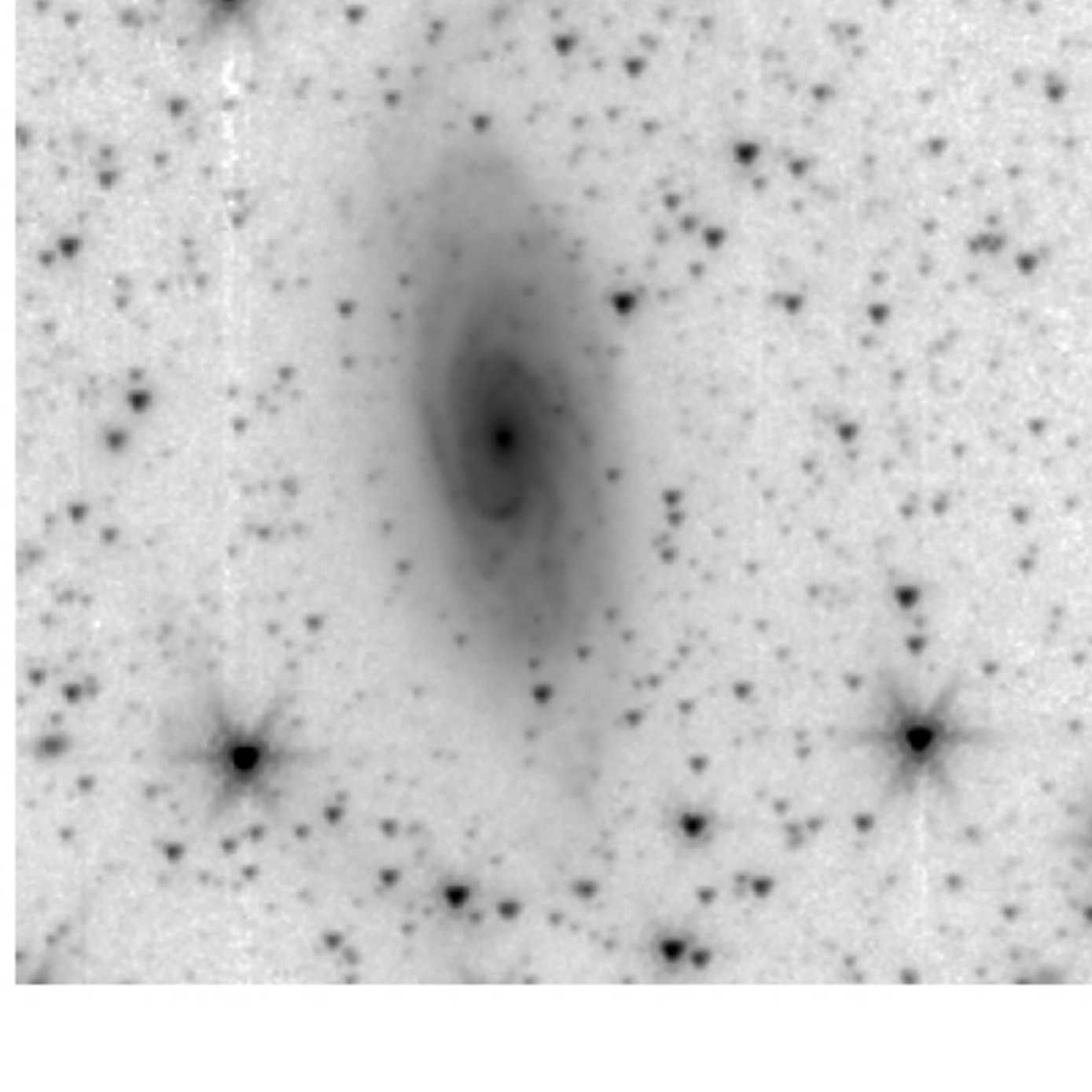}&
  \includegraphics[width=4cm]{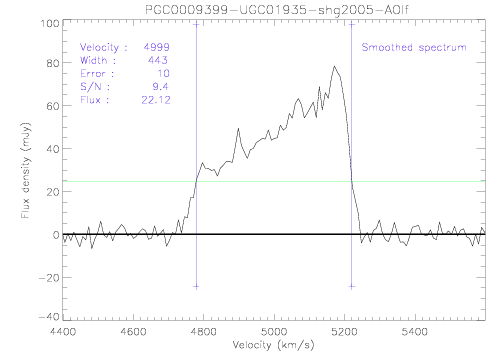}&
 \includegraphics[width=3cm]{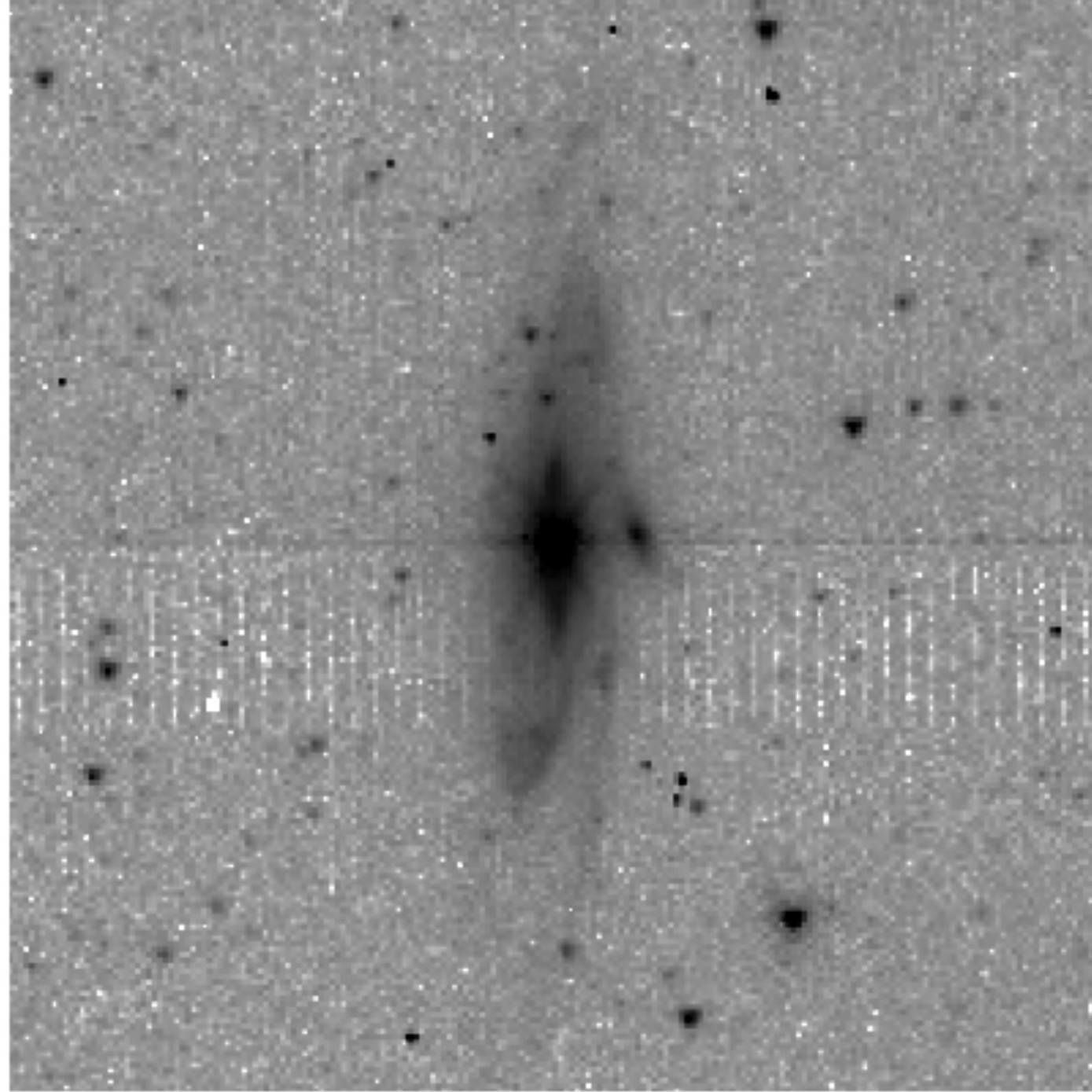}\\
 Parkes 64m && Arecibo\\
   \hline  
     \hline  
PGC18089	& 	2.29 / 5791  ~$km\,s^{-1}$  &    
PGC28196   & 	 2.27 / 4772 ~$km\,s^{-1}$ \\
 \includegraphics[width=4cm]{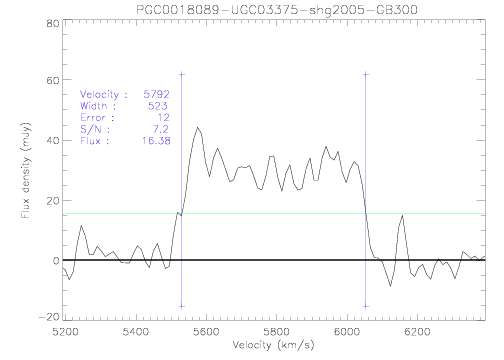}&
 \includegraphics[width=3cm]{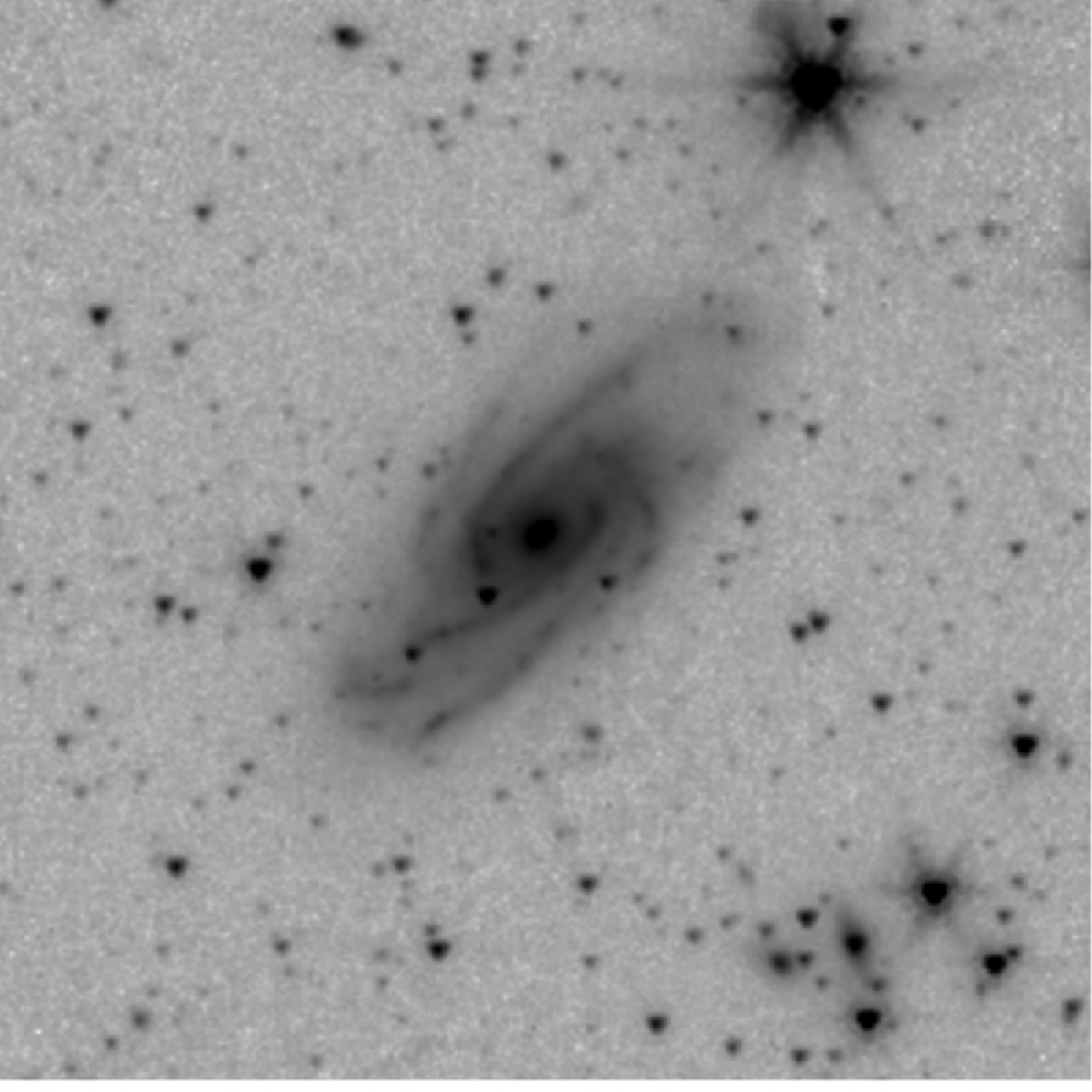}&
  \includegraphics[width=4cm]{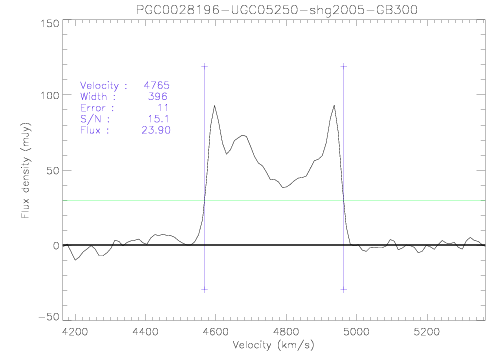}&
 \includegraphics[width=3cm]{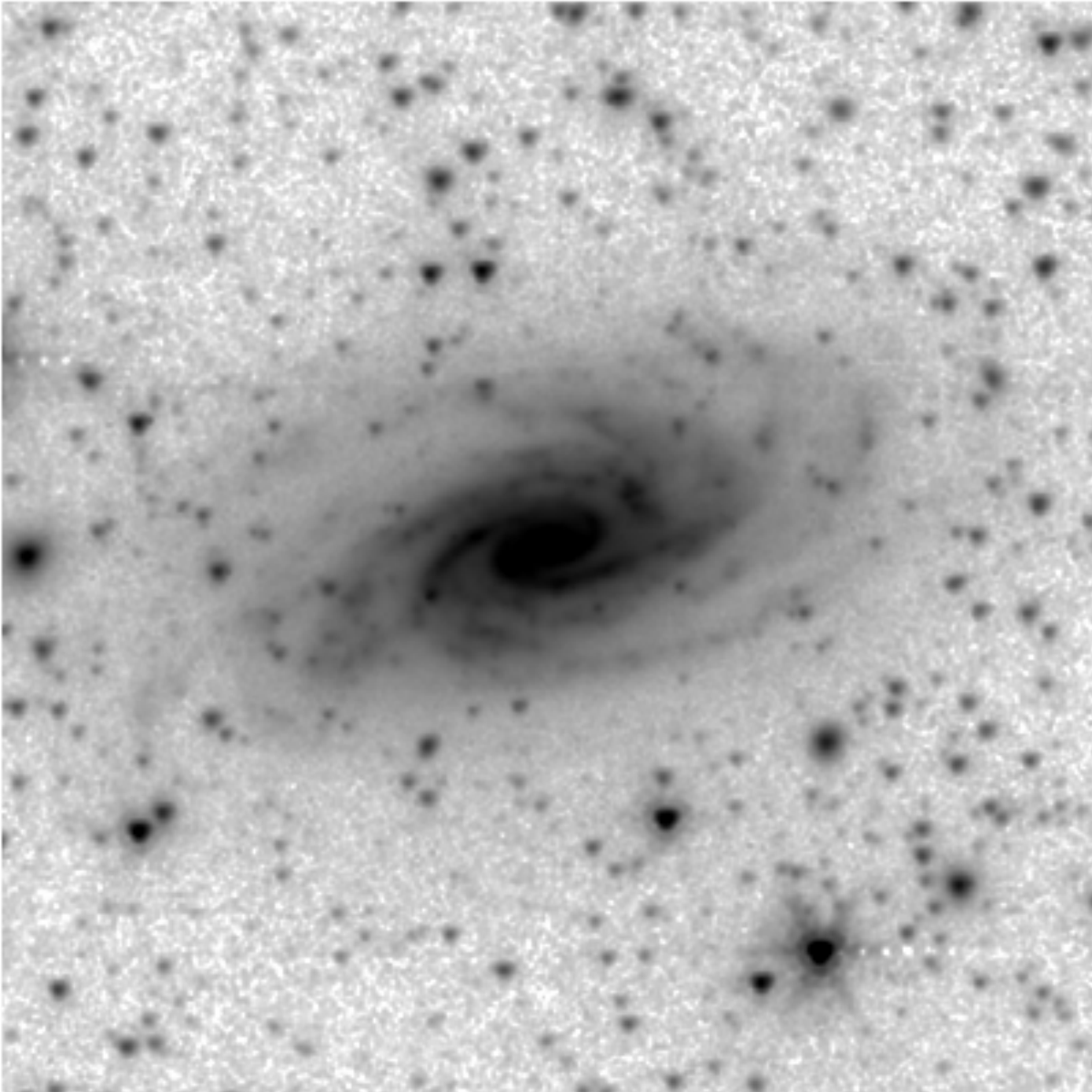}\\
GB300  && GB300 \\
   \hline  
\end{tabular}
\end{table*}

\clearpage


\bibliography{biblicomplete}

\end{document}